\documentclass[manuscript]{aastex}

\usepackage{amsmath}
\usepackage{multirow}
\usepackage{color}

\newcommand{\ionn}[2]{#1\,{\sc #2}}
\newcommand{\ionnii}{\ionn{N}{ii} $\lambda$1085}
\newcommand{\ionciv}{\ionn{C}{iv} $\lambda\lambda$1548,1551}				%
\newcommand{\ionnv}{\ionn{N}{v} $\lambda\lambda$1238,1242}

\newcommand{\ionskivb}{\ionn{Si}{iv} $\lambda\lambda$1393,1402}
\newcommand{\ionciii}{\ionn{C}{iii} $\lambda$1175}
\newcommand{\ionskiiia}{\ionn{Si}{iii} $\lambda$1206}
\newcommand{\ionskiiib}{\ionn{Si}{iii} $\lambda\lambda$1298,1304}

\newcommand{\ioncii}{\ionn{C}{ii} $\lambda$1335}

\newcommand{\sci}[2]{#1$\times$10$^{#2}$}

\newcommand{\msun}{M$_\odot$}
\newcommand{\rsun}{R$_\odot$}
\newcommand{\sunyr}{\msun\ yr$^{-1}$}
\newcommand{\massrate}[2]{#1$\times$10$^{#2}$ \sunyr}
\newcommand{\mwd}{\hbox{M$_{\hbox{\rm \scriptsize WD}}$}}
\newcommand{\rwd}{\hbox{R$_{\hbox{\rm \scriptsize WD}}$}}

\shorttitle{ACCRETION DISK WINDS}
\shortauthors{PUEBLA ET AL.}

\begin{document}

\title{A Method for the Study of Accretion Disk Emission in Cataclysmic Variables I: The Model}

\author{Ra\'{u}l E. Puebla and Marcos P. Diaz}
\affil{Instituto de Astronomia, Geof\'{\i}sica e Ci\^{e}ncias Atmosf\'{e}ricas,
    Universidade de S\~{a}o Paulo,\\
    R. de Mat\~{a}o, 1226, Cid. Universit\'{a}ria, cep 05508-090, S\~{a}o Paulo, SP, Brazil}
\email{raul@astro.iag.usp.br, marcos@astro.iag.usp.br}
\and
\author{D. John Hillier}
\affil{Department of Physics and Astronomy, University of Pittsburgh, \\
3941 O'Hara Street, Pittsburgh, PA, 15260, USA}
\email{hillier@pitt.edu}
\and
\author{Ivan Hubeny}
\affil{Department of Astronomy/Steward Observatory, The University of Arizona,\\ 933 N Cherry Ave., Tucson AZ 85721-0065, USA}
\email{hubeny@as.arizona.edu}

\begin{abstract}
We have developed a spectrum synthesis method for modeling the UV emission from the accretion disk from cataclysmic variables (CVs). The disk is separated into concentric rings, with an internal structure from the \citeauthor{wade98} disk-atmosphere models. For each ring, a wind atmosphere is calculated  in the co-moving frame with a vertical velocity structure obtained from a solution of the Euler equation.  Using simple assumptions, regarding rotation and the wind streamlines, these 1D models are combined into a single 2.5D model for which we compute synthetic spectra. We find that the resulting line and continuum behavior as a function of the orbital inclination is consistent with the observations, and verify that the accretion rate affects the wind temperature, leading to corresponding trends in the intensity of UV lines. In general, we also find that the primary mass has a strong effect on the P-Cygni absorption profiles, the synthetic emission line profiles are strongly sensitive to the wind temperature structure, and an increase in the mass loss rate enhances the resonance line intensities. Synthetic spectra were compared with UV data for two high orbital inclination nova-like CVs --- RW Tri and V347 Pup. We needed to include disk regions with arbitrary enhanced mass loss to reproduce reasonably well widths and line profiles. This fact and a lack of flux in some high ionization lines may be the signature of the presence of density enhanced regions in the wind, or alternatively, may result from inadequacies in some of our simplifying assumptions.
\end{abstract}


\section{INTRODUCTION}\label{intro}

Cataclysmic Variables (CVs) are semi-detached binary systems in which a Roche-lobe filling main sequence star (secondary) transfers matter to a white dwarf (primary).  In the case of non-magnetic systems the mass transfer is made through an accretion disk. In Nova-like and dwarf novae in outburst, the disk is the main source of radiation. The inner disk emission dominates the ultraviolet (UV) spectrum \citep{warner95}.

Several efforts have been made to understand the spectral features of accretion disks within different physical frameworks, from simple composite black-bodies  \citep{lynden69,lynden74,tylenda77,pringle81} to more realistic atmosphere-disk emission models that take into account line blanketing, Doppler broadening and limb darkening \citep{dous89,diaz96,wade98}. The influence of the white dwarf (WD), disk rim and orbital phase on the spectrum have  also been studied \citep{linnell96,linnell07a}. These attempts to reproduce UV continuum features have been widely tested with UV data \citep[e.g.][]{long94,diaz99,nadalin01,engle05,linnell08} and there have been some successes in reproducing some photospheric spectral features,  for example the absorption profiles of \ionskiiib\ and \ioncii, and the light curve through eclipse. However, these models have deficiencies  --- they are unable to reproduce the flux level and continuum color at the same time \citep[e.g.][]{wade88,puebla07}. Another significant problem with these models is that even though they reproduce some photosphere spectral characteristics and some absorption profiles, they fail to reproduce the optical and UV emission lines that are observed for CVs. 

Since the pioneering studies to model the emission lines from dwarf novae in quiescence \citep{tylenda81,williams80,williams82,williams88}, new physical scenarios have been proposed to explain the strength and the orbital behavior of emission lines. The P~Cygni profiles of \ionciv, \ionskivb\ and \ionnv\ observed in low to intermediate orbital inclination systems in high state, provide convincing evidence of mass loss through strongly accelerated winds (terminal velocities ranging from 2000 to 5000 km s$^{-1}$). The observed profiles resemble UV spectral features observed in the OB stars, and this has led to the idea that these outflows could also be line-driven winds \citep{cordova82,cordova85} driven off the accretion disk. However, some P~Cygni profiles in CVs show differences from those observed in OB stars. For example, the deepest flux level in the blueshifted absorption component is closer to the transition wavelength, while for hot stars it is close to the wind terminal velocity \citep[e.g.][ and references therein]{shlosman93,prinja95}. In addition, a correlation between disk wind signatures and the accretion rate has been found \citep{puebla07}. For systems with low accretion rate (e.g., dwarf novae in quiescence) the absorption component is absent or indiscernible in an asymmetric emission profile, while for high accretion rate systems, a deeper blueshifted absorption component is often seen. 

Studies on eclipsing systems indicate an axi-symmetric wind instead of spherical geometry and a highly stratified ionization structure of the emission line region in CVs  \citep{king83,drew87,mauche94,mason95}. The blueshifted absorption component gradually weakens as the orbital inclination increases. Thus, for high inclination systems the spectrum is mainly dominated by quasi-symmetric emission lines.

\cite{mason95} showed variations on the line profile and intensity through the eclipse in UX UMa, and  concluded that the lines are less eclipsed than the continuum,  especially \ionciv. This suggests that the size of the emission region is comparable to that of the secondary star, and the line emitting region is larger than the continuum emitting region. 
Similar results were found for RW Tri and DQ Her \citep{cordova85}. More recently high-time resolution far-ultraviolet (FUV) observations of UX UMa reveal strong and blueshifted \ionn{C}{iii}, \ionn{N}{iii} and \ionn{O}{vi} absorption lines \citep[e.g.][]{froning03} that suggest active mass loss.

The first effort to understand disk-wind lines quantitatively was made by \cite{cordova82} who used pure stellar wind models \citep{olson82} to fit the deep absorption components of some lines from the dwarf nova TW Vir in outburst. These authors adopted a spherical velocity profile centered on the white dwarf, and estimated the mass loss rate as 10$^{-2}$ to 10$^{-3}$ of the accretion rate. Later, \cite{drew86} \citep[see also][]{drew85} studied the effect of an extended continuum source on line profiles. They pointed out the strong influence of the mass loss rate on the line strength and the weak effect on the blue wing absorption. These authors also found that the more dense the wind, the bluer the P~Cygni absorption minimum. Finally, they applied their theory to IUE observations of RW Tri and UX UMa, and calculated a mass loss rate $\gtrsim$10$^{-10}$ M$_\odot$ yr$^{-1}$.

A second generation of models was developed by \cite{shlosman93} (hereafter SV93). To generate profiles for \ionn{C}{iv}, they developed a kinematical model with a bi-conical geometry, and with a stellar velocity law along tilted stream lines. An extended blackbody continuum source was constructed from a \cite{shakura73} flat disk, the white dwarf, and the boundary layer (BL). These authors assumed an isothermal wind with an ionization structure calculated through the UV photoionization-dominated media (H, He and CNO) and used the Sobolev approximation for radiative transfer. They found a series of profiles that show the rotation effects on line shape, and the strong influence of wind geometry \citep{vitello93}. They estimated the mass loss rate to be within $\sim$4\% to 15\% of the accretion rate. Later, they showed the impact of wind rotation on the profile for high orbital inclination systems such as V347 Pup \citep{shlosman96}.

Another effort to understand the wind properties through line synthesis was made using Monte Carlo radiative transfer methods.  \cite{knigge95} assumed a stellar wind velocity law, an isothermal gas, and that the line is formed by pure scattering. These methods helped to analyze the profile structure for a single line in more detail. After that, Monte Carlo methods were also used to calculate the ion and temperature structure of a bi-polar wind using the Sobolev approximation. This was the first effort to model the full CV spectrum in a broader wavelength range \citep{long02}. However, \cite{long06} pointed out the difficulties in reproducing more than one observed line with the same wind parameters. More recently, this method was improved \citep{sim05} allowing the formation of recombination lines and applied to RW Tri and UX UMa HST UV data \citep{noebauer10}. These models yield interesting and encouraging results about line profiles and disk wind structure and their results are compared with ours bellow. Particularly, previous models do not self-consistently compute the disk-wind interface. In this work we attempt to do it.

The aim of this work is to better understand the ionization structure, geometry and physical parameters of the line emitting region in CVs. To achieve this goal we have used disk-wind models, that include detailed wind physics (\S \ref{method}), to simultaneously model several line features observed in the UV spectra of CVs. The dependence of synthetic spectra on parameters such as the white dwarf mass (\mwd), accretion rate (\.M$_a$), mass-loss rate (\.M$_w$), orbital inclination ($i$), and wind geometry was investigated. A description of the method, including the calculation of the velocity laws, wind structure and spectrum synthesis is presented in \S \ref{method}. Extensive simulations are shown in \S \ref{models}. A comparison of synthetic spectra with UV and FUV data for the high inclination CVs RW Tri and V347 Pup is made in \S \ref{observations}.
A general discussion of our results is given in \S\ref{discussion}, while a brief summary and conclusions are presented in \S \ref{conclusions}.

\section{THE METHOD}\label{method}

\subsection{The Disk-Wind Model}

The observed UV line profiles of many high mass transfer CVs resemble those observed in O stars winds. Due to the similar spectral characteristics, it is feasible to suppose that in CVs the emission lines are formed in a wind ejected by the system and, similarly to stars, this outflow is driven by line radiation.

Actually, the disk wind is a complex 3D structure, but as was pointed out in section \ref{intro}, the wind seems to have an axi-symmetric geometry. From this fact, the models should use at least an 2.5D (2 spatial directions plus rotation) approximation, similarly 2.5D radiative models need to be used to predict the observed disk-wind spectra. However, a complete 2.5D hydrodynamic plus radiative transfer calculation would be much more complicated and time consuming than the 1D modeling. Since many basics of CV wind structure and acceleration are still not understood we have adopted an intermediate approach. We use 1D models to compute the atmospheric structure (including the deep layers of the disk, disk photosphere and wind) -- that is, the temperature structure, the ionization structure, and the level populations in vertical sense. We then combine a series of 1D models to create a 2.5D model, and use this 2.5D model to predict synthetic spectra. As we will show below, in this approximation the temperature in the extended wind region is controlled only by the radiation within the 1D model corresponding to the $r$ radius. This would not be desired, taking into account that the wind temperature far from the disk is controlled by the radiation from the whole disk. Despite these simplifying assumptions we got reasonable good UV line profiles when compared with observations (section \ref{observations}).  One advantage of our approach is that we correctly treat the photosphere-wind transition region as is described in section \ref{atm}.

\begin{figure}[!t]
\centering
\leavevmode
\includegraphics[width=1.0\columnwidth]{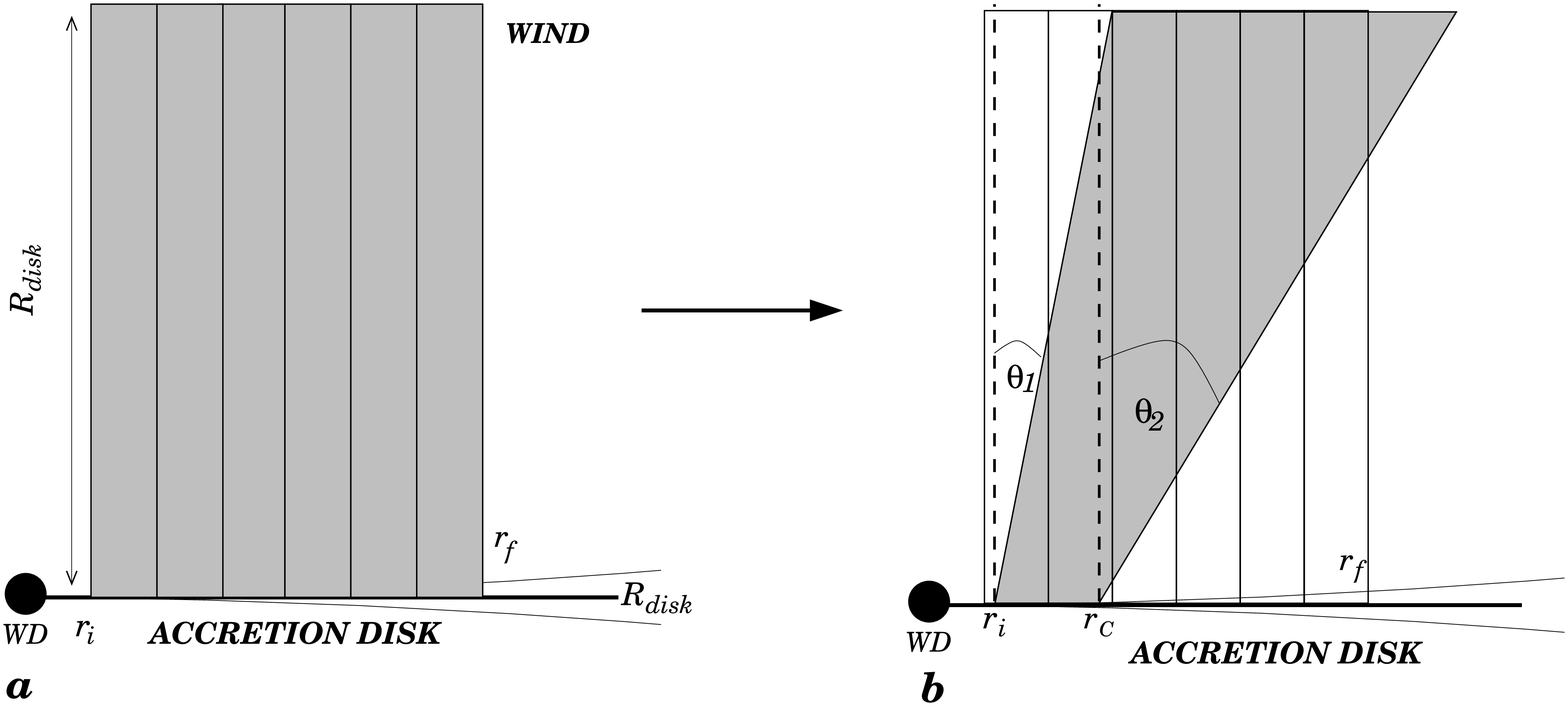}
\caption{Geometry of system used to model the disk wind and to calculate the spectrum synthesis. The geometrical parameters for spectral synthesis are thus defined (see text).}
\label{fggeometry}
\end{figure}

The model disk atmospheres and synthetic spectra are calculated using an arrangement of concentric wind atmospheres. The base of each atmosphere is located at the disk plane at a distance ``$r$'' from its  axis of symmetry. Each wind atmosphere has its own internal disk structure, disk photosphere and vertical wind. The atmospheres are placed side by side, thus forming the 2D structure of disk, photosphere and wind. The atmosphere structure calculus requires a vertical velocity field, which is obtained consistently with the physical conditions in the disk, such as gravity, radiation and density for a wind within \cite{castor75} (hereafter CAK) framework, as explained in section \ref{velaw}. Once the atmosphere structures are calculated the opacity and emissivity in the co-moving frame (CMF) are computed for each depth in the vertical grid. These values are then stored on a two-dimensional (and finer $r,z$) 2D grid through interpolation. This grid covers from an inner radius $r_i$, to an outer radius $r_f$ as it is shown in figure \ref{fggeometry}a. The vertical size of the wind is set equal to $R_{disk}$. The region between $r_i$ and $r_f$ embraces the radii where the wind is expelled. Each of wind atmosphere is calculated in the vertical sense using the method described in section \ref{atm} that is widely used in the of study hot star winds. We obtain thus, a cylindrical wind with a vertical velocity field.

\begin{figure*}[!t]
\centering
\leavevmode
\includegraphics[width=0.4\columnwidth]{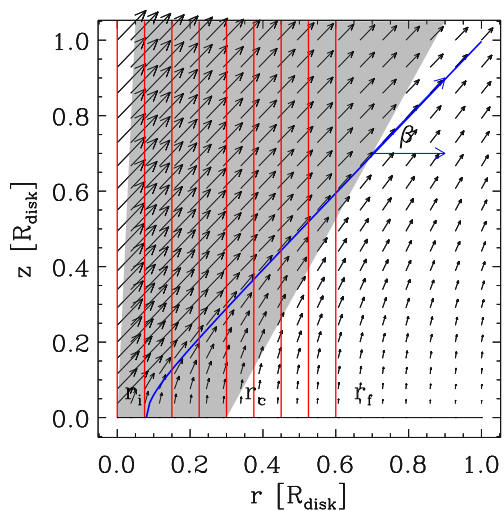}
\caption{Combined V$_r$ and V$_z$ velocity fields. Red lines show the radial assembly of wind atmospheres. Blue line shows a hyperbolic stream line with an asymptotic aperture angle $\beta$.}
\label{fgvelfield}
\end{figure*}

It is necessary to introduce two more components to velocity field, a radial component $V_r$ and azimuthal component $V_\phi$, that together with $V_z$ will bear a 3D velocity field. We arbitrarily set the $V_r$ component in order that the projection streamline describes an hyperbolic trajectory with an asymptotic angle $\beta$ (see figure \ref{fgvelfield}). The streamlines don't follow 1D vertical models, their bases are on a point of the disk plane with radius $r_o$ corresponding to an hyperbolic trajectory that passes through the ($r,z$) point with $V_z$ and $V_r$. The $V_\phi$ component was calculated in order to conserve the specific angular momentum from the wind base (keplerian) and to simulate the divergence of the wind streamlines. The base stream radius $r_o$ was calculated from the trajectory equation for each ($r,z$) point:
\begin{equation}
z^2=\tan^2\beta (r^2-r^2_o)
\end{equation}
Thus the velocity field is given by
\begin{eqnarray}
V_r(r,z)=
\begin{cases}
 \frac{z}{r \tan(\beta)}V_z(r,z),     & \text{if $z<r\tan(\beta)$},\\	
 \frac{1}{\tan(\beta)}V_z(r,z) & \text{if $z\geq r\tan(\beta)$}
\end{cases}
\end{eqnarray}

\noindent
and the specific angular momentum conservation ($r_o\sqrt{G\mwd/r_o}=rV_\phi(r,z)$) leads to:

\begin{eqnarray}
 V_\phi(r,z)=
 \begin{cases}
  \sqrt{\left( 1-\frac{z^2}{r^2\tan^2(\beta)}\right)^{1/2}\frac{GM}{r} } & \text{if $z<r\tan(\beta)$},\\
  V_\phi(r,h) & \text{if $z\geq r\tan(\beta)$}
  \end{cases}
\end{eqnarray}  

\noindent
where $V_r$ is the radial velocity, $V_\phi$ is the azimuthal velocity, $\beta$ is the asymptotic aperture angle of the streamlines (figure \ref{fgvelfield}) and $V_z$ is the velocity law calculated as shown in \S \ref{velaw}. The parameter $h$ is the corresponding value of $z$ of the last vertical grid point such that $V_\phi$$>$0. This is to avoid discontinuity beyond $z$=$r\tan(\beta)$. In this work we use $\beta$=45$^\circ$. Such a value correspond to an average of the values found in \cite{proga99} and \cite{pereyra03}. It was kept fixed as we needed to constrain the number of free parameters to explore. 

In the spectral synthesis we have introduced additional parameters to modify this rigid geometry in order to explore their influence on the spectral features. The figure \ref{fggeometry}b shows those parameters and their meaning. The aperture angles $\theta_1$ and $\theta_2$ limit the regions which will be taken into account in the final radiative transfer calculus. There is no matter within the cone defined by $\theta_1$ and outside the cone defined by $\theta_2$. In the last case, the region inside the cone defined by $\theta_2$ and beyond $r_f$ is filled by the same structure corresponding to the last wind atmosphere calculated on $r_f$. The $r_c$ radius also limits the regions that eject wind, but keeping the wind structures considered above the line defined by $\theta_2$. There is not a consistent reason for these approximations, we introduce them in order to make our analysis more flexible and to mimic the well studied geometry of disk winds \citep[e.g.][]{shlosman93,long02}. Using those parameters we were able to follow the geometry predicted in hydrodynamic models \citep[e.g.][]{proga98,pereyra00,pereyra03}.

It is clear that for regions close to the symmetry axis the corresponding streamlines wouldn't hit the disk. Those streamlines will be empty (or filled) depending on the $\theta_1$ value, this is parametrized to simulate the kinematics of the wind in these regions without solving the wind dynamics in 2.5D. Also, from this approximation the mass is not exactly conserved trough the wind streamlines, specially in the regions far from the disk photosphere. Despite of this, we can calculate an approximate value for the total mass-loss rate, \.{M}$_{w}$ by: 

\begin{equation}
\dot{M}_w= 4 \pi  \int_{r_i}^{r_c}  \dot{m}(r) rdr 
\end{equation}

\noindent
where $\dot{m}(r)$ is the mass loss flux from disk that is calculated from Euler equation as it is shown below (section \ref{velaw}). An additional factor of 2 has been included to account for both sides of the disk.  In our approximation the calculated mass loss value will be slightly different than the “real” value that would produce the line emission. This due to the influence of $\theta_1$, $\theta_2$, and $r_C$ on the volume size that is taken into account in the spectral synthesis. The more these parameters are away from a cylindrical model, the farther this value of mass loss will be off. However, an increase in $\theta_1$ is roughly compensated by an increase in $\theta_2$. Considering the range of parameters used in this work, we can estimate that in worst cases the departure can reach $\sim$20 \%.

Using the resulting emissivity and opacity values, mapped into the observer's frame, the radiative equation is solved exactly along rays that are defined by a set of impact parameters distributed concentrically in the disk plane. By integrating outgoing specific intensities over the solid angle comprised of the disk plus wind we obtain the synthetic spectrum. The details of each step are described in the following sections.

\subsubsection{The vertical velocity laws}\label{velaw}

The velocity law is required for defining the density profile and radiative transfer in the disk wind. It is calculated by solving the Euler equation for a vertical wind with the conditions found at each ring location in the accretion disk. It is important to note that unlike stellar atmospheres the gravity at a given radius increases with height inside the atmosphere up to a maximum at $\frac{z}{r}$$\sim$0.7. Gravity then decreases at greater heights, following a $\frac{1}{z^2}$ law at large distances above or below the disk. The flow base is at the disk plane and extends out to a distance equal to the disk radius. Here, we solved the momentum equation in the context presented by \cite{pereyra04}. This approximation has already been used to study hydrodynamic properties of disk winds in previous works \citep[e.g.][]{vitello88,pereyra97,pereyra05}. The Euler equation for an isothermal wind can be expressed as follows \citep{pereyra05}:

\begin{equation}\label{eqeuler}
\left( 1-\frac{s}{\omega} \right)\frac{d\omega}{dx}=-ga+fa\left( \frac{1}{\dot{m}}\frac{d\omega}{dx}\right)^{\alpha}+ \frac{2s}{a}\frac{da}{dx} ,
\end{equation}

\noindent
and following the \cite{pereyra04} and \cite{pereyra05} reasoning for a case of a vertical disk-wind we found that:

\begin{eqnarray}\label{eqeuler2}
 w &=& \frac{V^2_z}{v_o^2}\nonumber \\
 x &=& \arctan \left(z/r\right)\nonumber \\
 a &=& \sec^2(x) \nonumber \\
 s &=& \frac{b^2}{v_o^2} \\
 \dot{m}&=&\dot{m}(r)/\dot{m}_o \nonumber \\
 \dot{m}_o &=&\alpha\left(1-\alpha\right)^{(1-\alpha)/\alpha}\gamma_o^{1/\alpha}\left(\frac{2r}{v_o^2}\right)^{(1-\alpha)/\alpha} \nonumber\\
 \gamma_o&=&\frac{\kappa^{ref}_e \sigma T^4(r)k}{c}\left(\frac{1}{\kappa^{ref}_e v_{th}}\right)^\alpha	 \nonumber \\
 f&=&\frac{1}{\alpha^\alpha(1-\alpha)^{1-\alpha}}\zeta(x) \nonumber \\
  g &=& \sin(x)\cos^2(x)- \frac{\kappa^{ref}_e \sigma T^4(r)r^2}{cG\mwd}\cos^2(x) \nonumber  \nonumber
\end{eqnarray}

\noindent
In the above, the effective temperature used in this work follows the standard accretion disk model radial profile:

\begin{equation}\label{eqtemp}
	T(r)=\left( {3G\mwd\dot{M}_a}\over{8\pi \sigma \rwd^3}\right)^{1/4}\left(
	 {\rwd}\over{r}\right)^{3/4}\left[ 1-\left( 
	{\rwd}\over{r}\right)^{1/2}\right]^{1/4} ,
\end{equation}

\null
\noindent
\citep[see][]{shakura73,lynden74} where $r$ is the distance from the disk center to respective annulus measured at disk plane, \rwd\ is the compact star radius, $\sigma$ is the Stefan-Boltzmann constant, $G$ is the gravitational constant, \.{M}$_a$ is the mass accretion rate and \mwd\ is the WD mass. Assuming a 0 K pure carbon WD of \cite{hamada61}, $\rwd \propto \mwd^{\hbox{-0.7}}$ and the maximum disk temperature T$_{max}$ scales as $\sim$\mwd$^{\hbox{0.8}}$. Furthermore, $v$ is the velocity along a streamline that begins on the disk, $a$ is the geometrical factor that takes into account the streamline divergence \citep{vitello88}, $v_o$=$\sqrt{2G\mwd/r}$ is the escape velocity at the base of the wind, $b$=$\sqrt{k_BT(r)/\mu m_H}$ is the sound speed, $k_B$ is the Boltzmann constant and $\kappa^{ref}_e$=0.325 cm$^2$ gr$^{-1}$ is a reference value for the electron scattering opacity \citep{lamers99}. We have used an atomic weight $\mu$=1.28, similar to the solar value, and $m_H$ is the proton mass. The CAK parameters $\alpha$ and $k$ are physically related with the line opacity distribution and with the ratio of optically thick lines to optically thin lines considered in the line force. Thus for $\alpha$=1 only thick lines are taken into account, whereas for $\alpha$=0 only thin lines are considered. These parameters are not independent of each other as pointed out by \cite{gayley95} (see also CAK). The measured $\alpha$ and $k$ values (Table 1 in \cite{gayley95}) show a power-law distribution of lines and power law relation between these parameters. In this work we use this distribution for a range of $\alpha$ between 0.9 and 0.5. The function $g$ evaluates the gravity force corrected for electron scattering acceleration and normalized by $G\mwd/r^2$. The mass loss flux $\dot{m}(r)$ is normalized by $\dot{m}_o$, which in turn depends only on local disk characteristics. The parameter $\dot{m}_o$ is the equivalent to \.{M}$_{CAK}$ in \cite{pereyra04}.

The function $\zeta(x)$ represents the disk contribution to the radiation field, at a particular wind point. This radiation comes mainly from the disk surface and is used to calculate the line radiation force. Such a this force is normalized to the electron scattering force through the $\gamma_o$ parameter. In this work, we mainly used the analytical relation proposed by \cite{pereyra97} that takes into account the radiation contribution from the whole disk. Also, for some cases, we used the \cite{pereyra04} relation that prescribes $I$ models for inner regions and $O$ models for outer ones. Following \cite{pereyra04}, the $I$ models try to mimic the vertical flux distribution of the inner standard disk model, and the $O$ models try to mimic the flux distribution of the outer regions. The former falls as $[1-(\frac{z}{2r})^2]^{-1}$ because most of radiation is coming from region immediately below. In the $O$ approximation the flux first grows with height as the wind see the inner (hotter) regions, then it decreases in the same sense as in the $I$ models. The match criterion for both models is the same terminal velocity. We designate $\zeta_1$ (eq. \ref{eqzeta1}) for \cite{pereyra97} parameterization and $\zeta_2$ for the one of \cite{pereyra04} (eq. \ref{eqzeta2}). A detailed analysis of equation \ref{eqeuler} behavior is presented by \cite{pereyra04} and \cite{pereyra05}.

\begin{eqnarray}\label{eqzeta1}
\zeta_1(x)&=&\sin^\alpha(x)\cos^2(x)\\
\nonumber \\
\zeta_2(x)&=& 
\begin{cases}\label{eqzeta2}
\frac{4}{4+\tan^2(x)} & \text{$I$ case} \\
\frac{1+2\tan(x)}{1+\tan(x)}\cos^2(x) & \text{$O$ case}
\end{cases}
\end{eqnarray}

\null
\noindent
In this work we iteratively solve the Euler equation to find the critical point and the mass loss corresponding to the critical solution (CAK). The boundary conditions for each iteration were set by calculating an atmosphere disk model \citep{wade98} and a velocity law that follows its density structure. This velocity law is computed from the disk plane to an outer optical depth $\tau_R$=10$^{-3}$ (where $\tau_R$ is the Rosseland optical depth). Then we find the height $z_c$ where $\tau_R(z_c)$=1 and the corresponding vertical velocity $v_c$ and acceleration $v'_c$ at this point. Equation \ref{eqeuler} is solved with the boundary conditions $v(z_c)=v_c$ and $\frac{dv}{dz}(z_c)=v'_c$. The theory of stellar atmosphere winds shows that the Euler equation has two solutions, the one of them, the \emph{lower branch}, goes from subsonic region to the $Parker$ point (in our case where $h=ga- \frac{2s}{a}\frac{da}{dx}=0$). The \emph{upper branch} solution goes from infinity to the sonic point (where $V_z=b$). Thus the region where the two solutions co-exist is found between the sonic point and the $Parker$ point \citep{lamers99}. The only solution that can start subsonic and grow smoothly through the sonic point to infinity is the critical solution that connects the lower branch solution to the upper branch solution at the critical point (CAK). The lower branch solution, even supersonic, fails when it reaches the $Parker$ point, but in our simulations this point has never been reached. That is the reason why we adopted the lower branch solution in our spectrum synthesis. These solutions bear lower terminal velocities that are more compatible with the observations. The critical solutions are found using the criteria exposed by \cite{pereyra05} for the existence and continuity.

\begin{figure}[!t]
\centering
\leavevmode
\includegraphics[width=0.5\columnwidth]{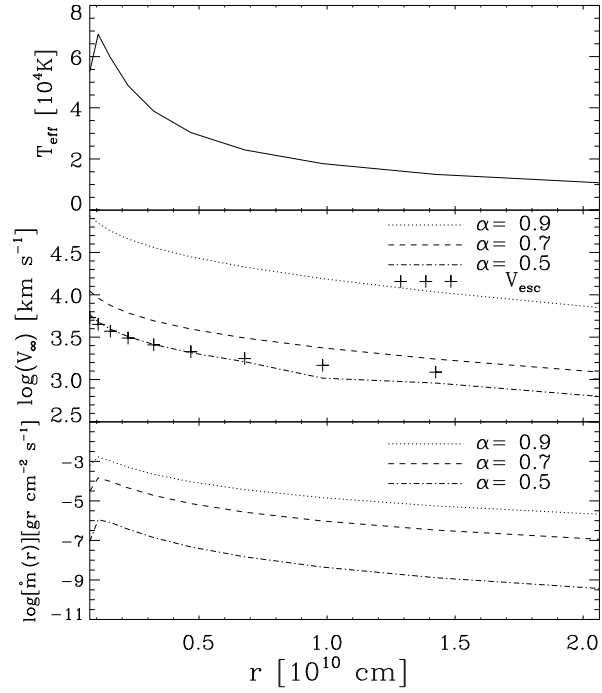}
\caption{Euler equation solutions for a disk wind with \mwd=1 M$_\odot$ and \.{M}$_a$=10$^{-8}$ M$_\odot$ yr$^{-1}$. The upper panel shows the temperature distribution. The middle panel shows the terminal velocities for the upper branch solutions for the three $\alpha$ (0.9, 0.7 and 0.5) values compared with escape velocity. The bottom panel shows mass loss fluxes that result from critical solutions.}
\label{fgvelmodels}
\end{figure}

The dependence of the terminal velocities on $\alpha$, as well as the dependence of the mass loss flux, are shown in Figure~\ref{fgvelmodels}. These values were calculated using the ``$\zeta_2$'' function with \mwd$=0.8$ and M$_\odot$, \.{M}$_a$=1$\times$10$^{-8}$ M$_\odot$ yr$^{-1}$. It was found that the critical point position is strongly dependent on $\alpha$, as are the terminal velocities. The mass loss flux also is dependent on $\alpha$ and $k$, as evident from Equations \ref{eqeuler} and \ref{eqeuler2}; it has also a close relation with the effective temperature at the base of each wind model. This relation is similar to that observed in hot stars, as well as the relation between the total mass loss rate and luminosity. 

For our disk models, the mass-loss rates (integrating up to $r$=0.3 R$_\odot$=\sci{2}{10} cm) are 6.0$\times$10$^{-10}$ M$_\odot$ yr$^{-1}$, 5.0$\times$10$^{-11}$ M$_\odot$ yr$^{-1}$ and 3.4$\times$10$^{-13}$ M$_\odot$ yr$^{-1}$ for decreasing  $\alpha$ values from 0.9 to 0.5. These mass-loss rates are consistent with 2D and 3D hydrodynamic simulations \citep{pereyra97,pereyra03} and represent $\sim$ 0.04 \% to $\sim$ 10 \% of the accretion rate.

\subsection{Spectral Synthesis}\label{synthesis}

Once the wind model grid is calculated we are able to compute the CMF emissivity ($\eta_\nu$) and opacity ($\chi_\nu$) at each point. The opacities and source functions (S$_\nu$) in the CMF are interpolated into a finer grid in the $zr$ plane. Additionally, they are mapped into the observer's frame. In order to synthesize the spectrum, a new grid of log-spaced impact parameters on the plane of the disk is set from $p_{min}$=$1.05$\rwd\ to $p_{max}=R_{disk}[\tan(i)+1]$, where $i$ is the orbital inclination. In addition, a linearly spaced grid of azimuthal angles is set. A ray goes to the observer from each point of the 2D grid thus formed. Radiative transport through the inner disk and wind is calculated using points along each ray that are spaced 1 km s$^{-1}$ in projected velocity. For impact parameters greater than $r_{C}$ and lower than R$_{disk}$ the corresponding value of I$_\nu$($z=0$) is equal to the emissivity of a disk atmosphere at this radius as described by \cite{wade98}, taking into account the limb darkening \citep{diaz96}. Thus, for impact parameters $p_{min}<p<r_c$ we calculate the radiative transfer consistently from disk plane through the photosphere and wind. For impact parameters $r_c<p<R_{disk}$ the disk is treated as a surface that emits as a \cite{wade98} disk-atmosphere. Finally, for impact parameters $R_{disk}<p<p_{max}$, I$_\nu$($z=0$)=0.

The radiative transfer is calculated using a broad frequency grid in CMF and a frequency grid in the observer frame. For each point in the latter grid the radiation is calculated at each point along the ray. This is done through the corresponding observed frequency in CMF, using the shift caused by the velocity field projected in the observer direction at each point. The projected velocity is given by

\begin{equation}
 V_p=V_r\cos \phi \cos i + V_\phi \sin \phi \cos i + V_z\sin i \, .
\end{equation} 

The resultant specific intensities I$_\nu$ in the observer frame, that leave the system from each impact parameter are then interpolated into a finer frequency grid and also in a finer azimuthal grid. Those specific intensities are integrated over the solid angle of the disk plus wind to get the synthesized spectrum of the whole system (disk+wind). Our 2D code was adapted from an earlier 2D code by \cite{busche05}, used to compute spectra from rapidly rotating stars that have an expanding atmosphere. The main change involved a switch to polar coordinates and disk-like geometry, rather than the spherical coordinates used in the original code.

\subsubsection{The 1D Disk-Wind Atmosphere through CMFGEN}\label{atm}

The method used to calculate the 1D disk-wind structures is based on the CMFGEN code\footnote[1]{http://kookaburra.phyast.pitt.edu/hillier/web/CMFGEN.htm}. Several details this method are widely described by \cite{hillier98} \citep[see also][]{hillier90,hillier01,hillier03}. CMFGEN is used to solve simultaneously the statistical equilibrium equations (non-local thermal equilibrium or NLTE), the radiative transport equation in the CMF, and the radiative equilibrium equation. For simplicity (and feasibility) we assume that the perturbed radiation field depends only on the local populations (diagonal approach), or local plus neighboring depths (tridiagonal approach).  The latter has generally better convergence, but requires twice the memory. The only difference between these methods is the model convergence. Once the convergence is achieved, the results are the identical. This process is repeated until convergence of the populations is obtained. CMFGEN allows for line coupling, level dissolution, dielectric transitions and  electron scattering. In addition, charge exchange, Auger ionization, and the influence of X-rays can be treated --- these processes have not generally been included in previous accretion disk wind models \citep{hillier98},  although their influence on the line profiles has not been quantified yet.

The energy balance constraint, as expressed by the radiative equilibrium equation, is used for deriving the vertical temperature structure. Line blanketing plays a crucial role in determining the atmosphere structure and spectra and it is included. In the case of hot stars, line blanketing has a strong influence for $\lambda\lesssim$ 1300 \AA. A similar statement applies for accretion disks, especially in the inner regions which have similar effective temperatures.  Originally CMFGEN was written to treat spherically symmetric expanding outflows; this is not the case in accretion disk, whose geometry tends to be bipolar. CMFGEN was extended to allow the calculation of wind atmospheres with a plane-parallel geometry, as needed for thin disk models. In this work we use this geometry for each $cylinder$ (disk plus wind).

Even with a plane parallel geometry, CMFGEN still uses stellar-like parameters, namely,  the luminosity (in L$_\odot$), core radius (R$_p$ in 10$^{10}$ cm), external radius (in R$_p$ units), mass loss rate (\.M$_w$), besides the wind velocity law, atomic models, ionization states present in the gas, atomic species and their abundances. In the disk case we have different local physical conditions at each radius. These physical conditions are characterized by the effective temperature, mass loss rate and the height to which the wind extends. In order to get the correct input for CMFGEN a fiducial core radius R$_p$=10.051 is set for all models, while the external radius is set to the core radius plus disk radius, which is set equal to 0.8 R$_{L_1}$, where R$_{L_1}$ is the distance from the disk center to the inner Lagrangian point (figure \ref{fggeometry}a). This value is close to the truncation limit imposed by tidal forces \citep{osaki93}. Mass loss rates are obtained from the local mass loss fluxes, and luminosities are obtained from the stellar radius R$_p$ and the effective temperature. 

The wind atmosphere models are calculated taking into account the internal density structure of disk as is described by \cite{wade98} \citep[see also][]{hubeny90}. In these models the radiation field and physical parameters deep inside disk-atmospheres depend on the highly uncertain behavior of viscous dissipation, which is parameterized as a power law of the column density \citep[see][]{diaz96}. Once this structure is calculated, the corresponding vertical velocity field up to $\tau_{R}$=1 is computed using continuity equation for a thin disk, $\rho(z) v(z) = \dot{m}(r)$. This velocity field is matched with the Euler solution as was explained above. The structure calculus with CMFGEN thus, goes consistently from disk plane passing through the disk photosphere, wind-photosphere interface out to wind with high $z$ as high as R$_{disk}$.

The mass loss rate per unit area at the photosphere ($\dot{m}(r)$) is calculated at each radius as explained in the section \ref{velaw}. The wind atmospheres are not calculated out to the external disk radius. Instead, the maximum radius used to calculate the wind atmosphere depends on the effective temperature, and generally stops around $\sim$14000 K. The contributions to UV spectra and the influence on the main wind lines beyond these regions are small. Accretion disk photospheric emission \citep{wade98} is calculated for the outer regions and transfered through the inner when the rays shock it.

\section{MODEL RESULTS}\label{models}

In this section we analyze a set of simulations with the aim of studying the dependence of the UV spectrum on the model parameters. To do this, we calculated  the disk structure and wind structure, as explained in the previous section, in a limited parameter space which covers the observed one, as found in literature. A range of CAK parameters ($\alpha$,$k$) that bracket the values for hot stars is explored. For a given accretion disk model, $\alpha$ and $k$ define the mass-loss rate and the wind velocity law for each disk ring. Tables  \ref{tbmodels} and \ref{tbgeopar} show the parameters used in this work, with the physical parameters that characterize the wind listed in Table \ref{tbmodels}. It is important to note that the physical parameters are not independent, actually only the \mwd, \.{M}$_a$, $\alpha$, $k$ and $\zeta(x)$ are. The remaining are dependent, namely, T$_*$ depend on \mwd\ and \.{M}$_a$, as well as \.{M}$_w$ and V$_\infty$. In this work we have fixed the $\beta$ value and the $\zeta(x)$ function in order to not increase the number of free parameters. A brief discussion in the influence of the $\zeta(x)$ function on the ``photospheric'' line \ioncii\ is presented bellow (see section \ref{discussion}).

The primary mass (\mwd) and the mass accretion rate (\.{M}$_a$) determine the disk properties, which in turn act upon the wind temperature and wind ionization structure. The mass loss flux is not an independent parameter -- rather it is determined by the parameters describing the accretion disk, and the CAK parameters ($\alpha$,$k$)  through Equation \eqref{eqeuler}. 
We also varied  (see Table \ref{tbgeopar}) the angles $\theta_1$ and $\theta_2$ that limit/define the calculated wind region that is taken into account for the spectrum synthesis.  $\theta_2$ mainly determines the extended atmosphere size at the coolest wind atmosphere calculated (Fig. \ref{fggeometry}). The other geometrical parameters are the orbital inclination ($i$), the wind height (Z$_{max}$) and the disk radius (R$_{disk}$) that set the wind volume and the part of the wind hidden by the disk itself.

\subsection{Atmosphere structures}\label{structures}

\begin{figure}[!t]
\centering
\leavevmode
\includegraphics[width=0.5\columnwidth]{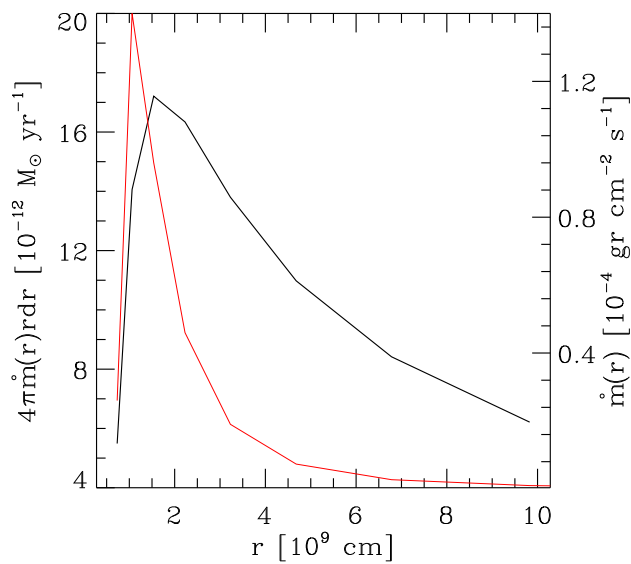}
\caption{Radial mass loss contribution in \sunyr (black  line) and mass loss flux distribution in gr s$^{-1}$ cm$^{-2}$ (red line). Both  correspond to model ``$e$'' (table \ref{tbmodele}).}
\label{fgmassflux}
\end{figure}

In this section we analyze the atmospheric structure in the vertical direction. The main factors that determine the ionization structure and the level populations are the electron temperature and density. As the structures of each disk-wind atmospheres are similar to each other, we chose one of the calculated models (model ``$e$'') in Table \ref{tbmodels} for the analysis --- differences between models will be highlighted, as necessary. Model ``$e$" has a white dwarf mass $\mwd =0.8$ M$_\sun$ and an accretion rate of  \.{M}$_a$=1.0$\times$10$^{-8}$ M$_\sun$ yr$^{-1}$ that yields a disk-temperature compatible with the most luminous CVs. The total mass loss-rate is \massrate{9.3}{-11}, which is 1\% of the mass accretion rate. 
While the mass fluxes vary by over two orders of magnitude, the outer regions of the disk (because of the weighting by area) make a very important contribution to the total mass-loss rate --- for model ``$e$'' each shell (i.e., the area between ring i, i+1) contributes roughly an equal  fraction [within a factor of 2] of the total mass-loss rate. The figure \ref{fgmassflux} shows this characteristic in a clearer manner for model  ``$e$''. While the flux mass falls with increasing radius, the total mass loss (quantified by $4\pi\dot{m}(r)rdr$) tends to stabilize at large radii. 

The physical parameters of each atmosphere are shown in Table \ref{tbmodele}. In this table the radius, in WD radius units, is shown for each ring, as well as the local mass loss flux, the vertical terminal velocity and the effective temperature following equation \eqref{eqtemp}. This table also shows the \textit{star-like} parameters used as input, namely: the stellar luminosity (L$_*$) and the total mass-loss rate (\.{M}$_v$).  

The last two atmospheres, corresponding to the largest and coldest rings, were not calculated. There are two reasons for this. The first one is  physical --- the contributions of these regions to the UV emission are low due to their low temperature. Further, these regions would only contribute to the line profiles if the wind density above the corresponding rings is high. The second reason is numerical ---  we found that, using plane-parallel geometry, the algorithms face convergence problems in cold, low $\log (g)$ and low density atmospheres. 

Detailed level populations in NLTE were considered for the H, He, C, N, O, Ne, Si and Fe. Their ionization states are shown in Table \ref{tbions} with the level and super-level numbers used for each ion.

\subsubsection{Temperature}

\begin{figure*}[!t]
\centering
\includegraphics[height=0.45\linewidth,width=0.45\linewidth]{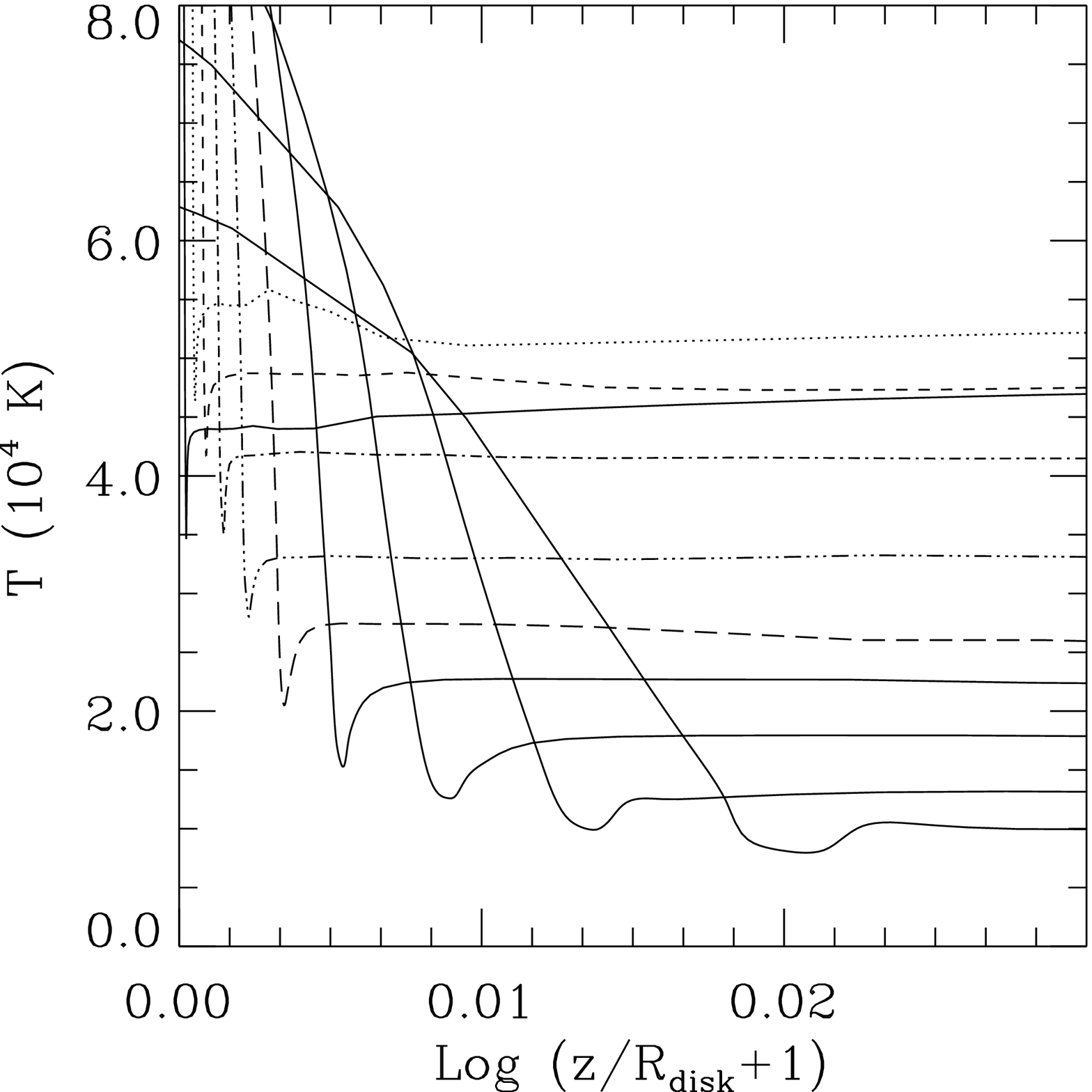}
\includegraphics[height=0.45\linewidth,width=0.45\linewidth]{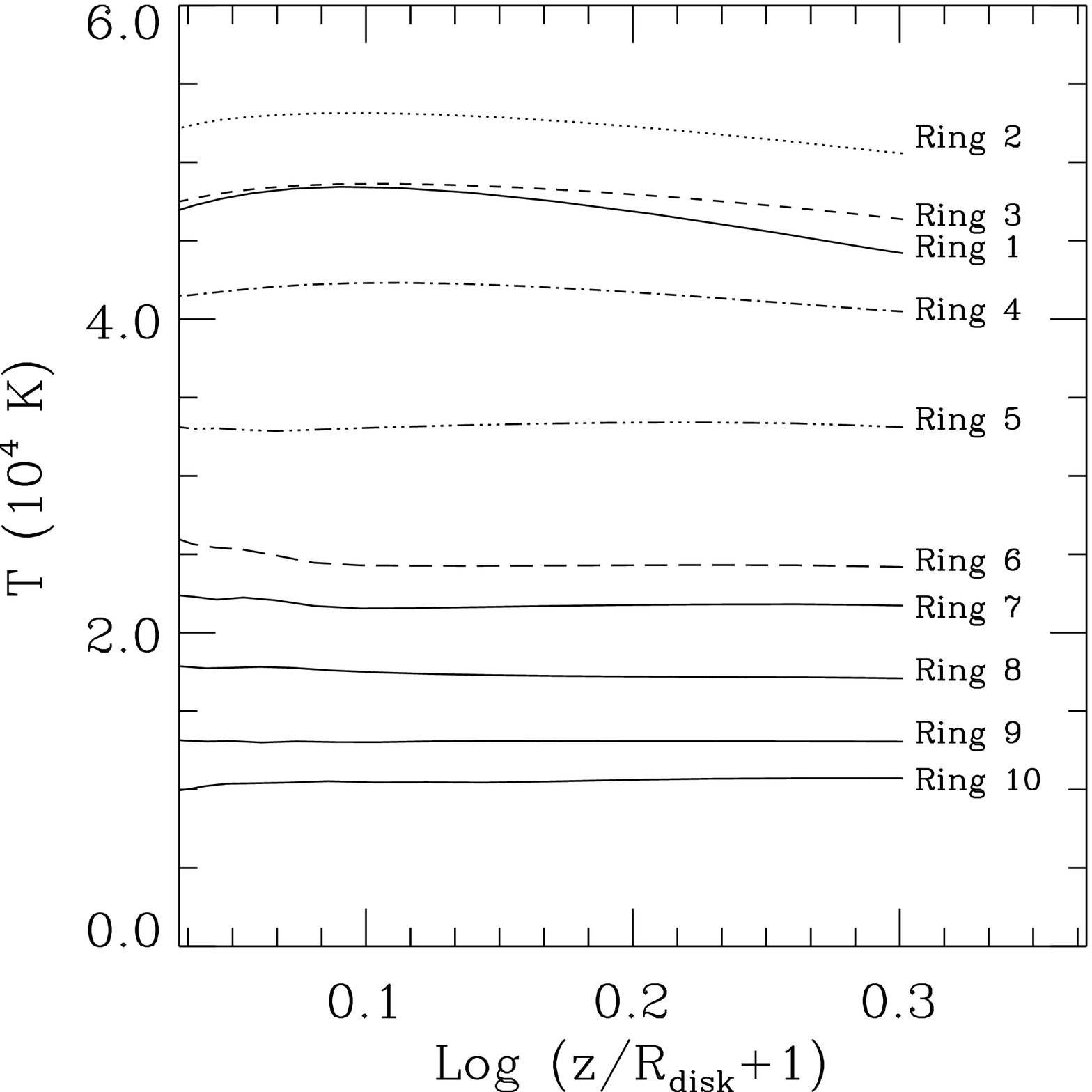}
\caption{Vertical temperature profiles for wind atmospheres. All rings in model ``$e$'' are shown.}\label{fgtemps}
\end{figure*}

In Figure \ref{fgtemps} the vertical electron temperature profiles are shown for each ring of  the disk plus wind system. In the left panel the temperature structure for the photosphere-wind transition region is shown. Every temperature structure for each ring displays the same morphology. From deep in the atmosphere the electron temperature decreases monotonically towards the photosphere, where the radiation begins to uncouple from matter. The temperature then rises because of an indirect effect of Lyman and Balmer lines on the Lyman-continuum heating until a value lower but close to the corresponding effective temperature is attained for each ring. This is a classical NLTE effect first discovered by \cite{auer69}. Note that the minimum temperature happens at a different height for each ring, as expected from the different height of $\tau_R$=1 for each disk photosphere structure. This is caused by the dependence of the  ``z'' component of  the WD gravity (which sets the atmospheric scale height in the disk) on $z$ and $r$ that generates the well known $disk$ $flaring$ \citep[][pag. 87]{frank02}. The right panel in Figure~\ref{fgtemps} shows the temperature behavior for extended wind regions for each atmosphere model. There the temperatures are almost constant with height, with a value close to the corresponding $T_{eff}(r)$. This behavior can be explained by the lack of radiative dilution in the plane-parallel approximation. Thus, the average outer electron temperature is T$_{e_\circ}$$\approx$$0.85T_{eff}(R)$. Actually the ratio T$_{e_\circ}$/$T_{eff}(R)$ is smaller for inner radii ($\sim$0.75), and larger for the outer zones where this ratio is almost equal to 1.

\subsubsection{Ionization Structure}

\begin{figure*}[!t]
\centering
\includegraphics[height=0.45\linewidth,width=0.45\linewidth]{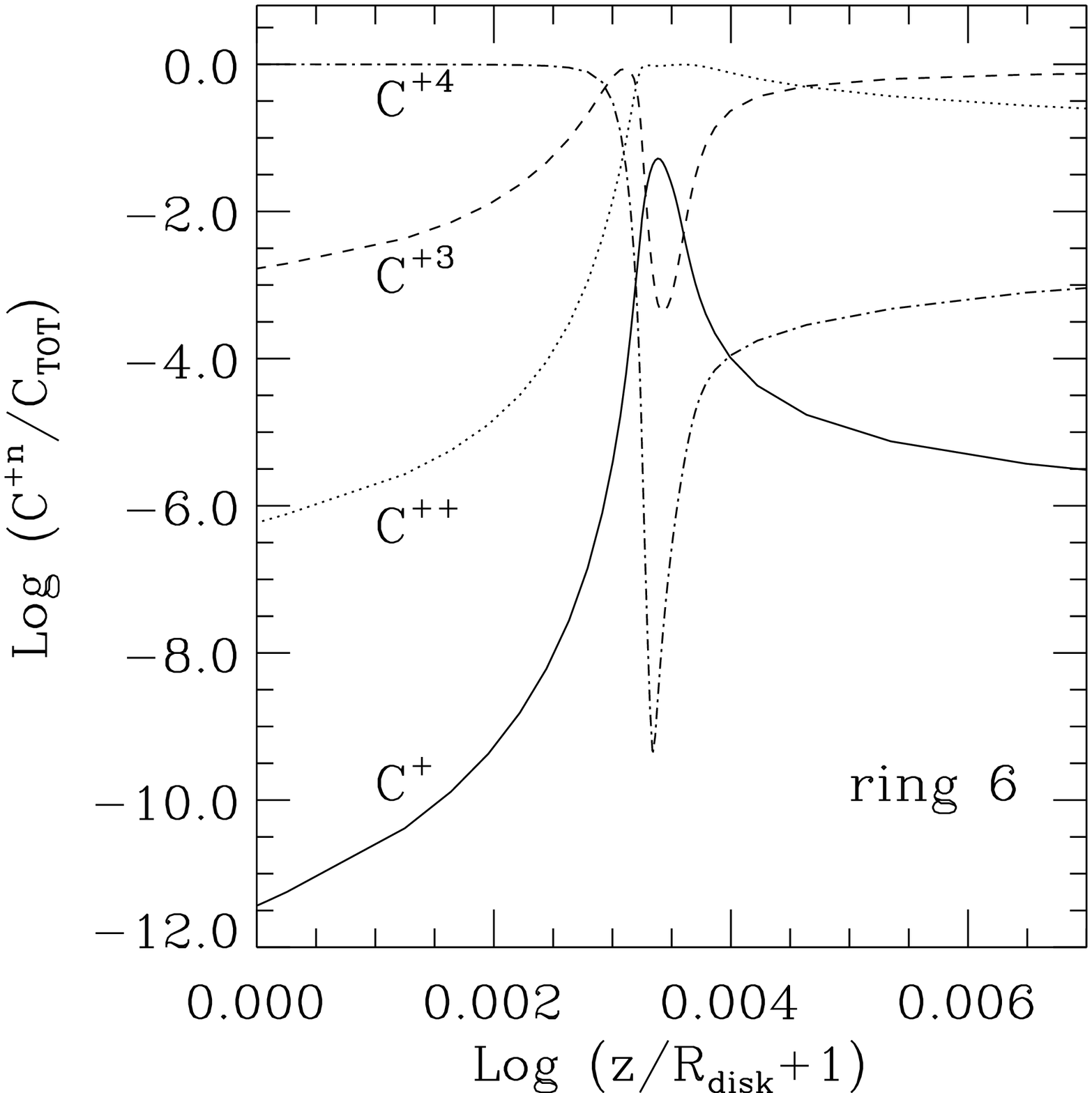}
\includegraphics[height=0.45\linewidth,width=0.45\linewidth]{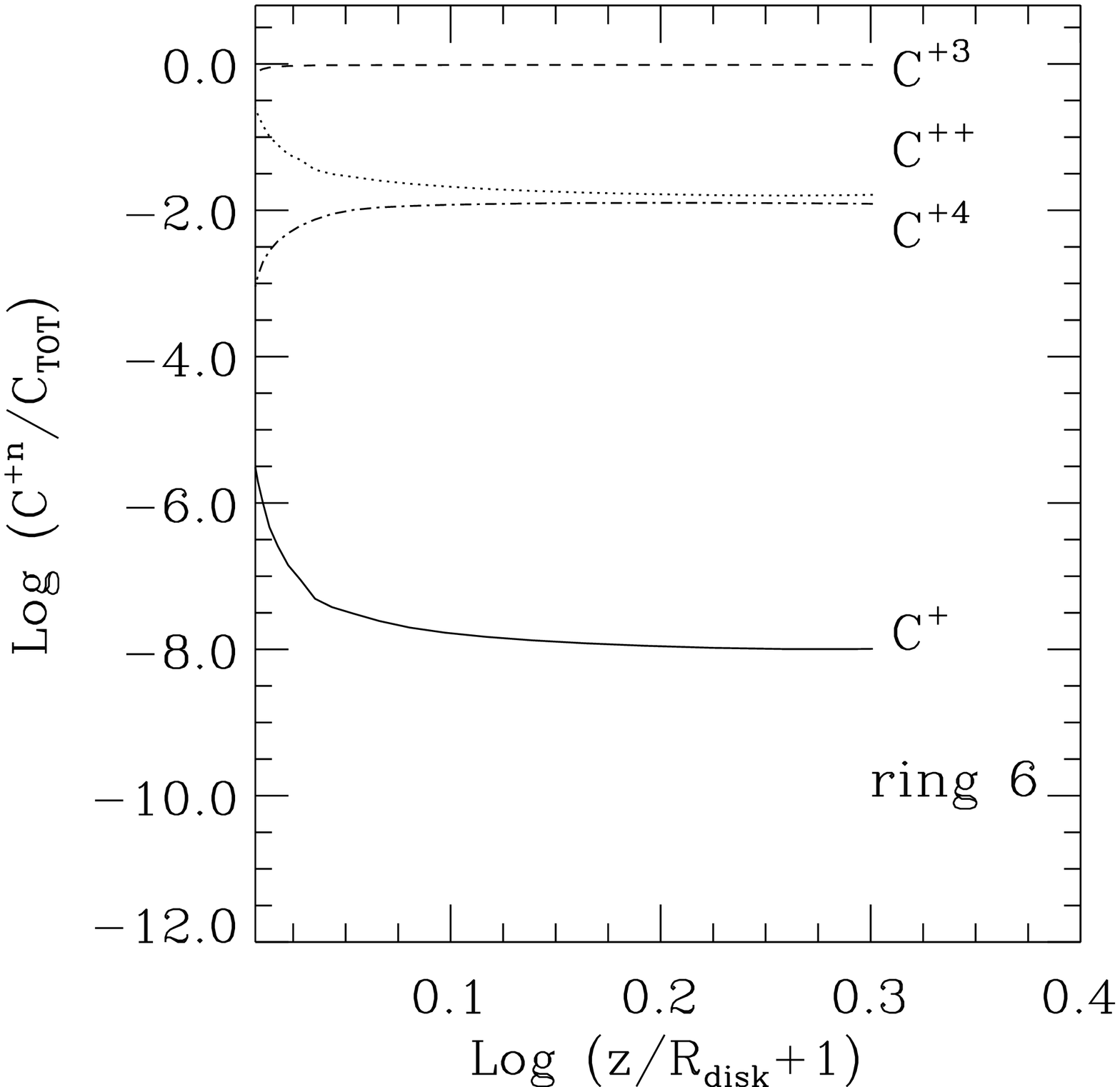}
\caption{Carbon vertical ionization structure for ring 6 of model ``$e$'' (T$_{eff}$=30300 K). The left panel shows the inner atmosphere, photosphere-wind transition. The right panel shows the extended wind region.}\label{fgionizing}
\end{figure*}

Due to the isothermal nature of the outer wind, and because there is very little dilution of the radiation field in the plane-parallel approximation, the ionization state for each atmosphere model is almost constant in the wind above the photosphere. However, strong variations in the ionization states are  found in the transition region between the photosphere and wind. These variations are stronger in cooler atmospheres. The ionization state for carbon, for ring 6 of model ``$e$'' which has an effective temperature of 30300 K, is shown in Figure~\ref{fgionizing}. Here, the left panel shows the photosphere-wind transition zone, and the right panel shows the extended wind region. The former indicates a highly structured ionization state, in particular above $\tau_R$=1. The dominant ion changes from C$^{+4}$ in the inner photosphere to C$^{+3}$, and from here to C$^{++}$. This happens in a small region with the electron temperature attaining its minimum at $\tau_R$$\sim$5$\times$10$^{-3}$. Immediately above this point, temperature begins to rise and the C$^{+3}$ concentration rises in the same way. When $\tau_R$$\sim$1.8$\times$10$^{-3}$ and the electron density N$_e$$\sim$3.7$\times$10$^{11}$ cm$^{-3}$, it becomes the dominant ion for the carbon. For greater heights the ionization structure tends to stabilize. That trend in the vertical direction is the same for all models.

As a consequence of the changes in effective temperature with radius, there are large radial changes in ionization. For instance, at the inner radii, the dominant ion for carbon is the C$^{+4}$ until the wind temperature is $\sim$39000 K at $r=4.6$~\rwd\ from the disk axis. For temperatures between 39000 K and 24000 K and $4.6\;\rwd<r< 9.7\;\rwd$ the dominant ion is C$^{+3}$. Beyond 9.7 \rwd, and for temperatures lower than  24000 K, the dominant ion is C$^{++}$ (out to  $r =29.6$ \rwd). Furthermore, if a high value for $\theta_2$ is used, the last model atmosphere is larger, and its ionization state becomes dominant in the wind. This behavior is followed by the other ions in the wind, depending on the radiation temperature that is coming from the model disk photosphere. In the same way, the level populations have strong variations in the photosphere-wind transition region, and a stable behavior in the wind zone. Beyond $\tau_R$=1, the level populations depart significantly from LTE. The departure coefficients increase with height up to a maximum, decreasing as gas enters the accelerated zone.

\subsection{Synthetic spectra}\label{spectra}

The effects of changing the model physical and geometrical parameters on the wind line profiles and continuum are described in this section. We have chosen models from Table~\ref{tbmodels} that differ in one physical or geometrical (Table \ref{tbgeopar}) parameter to guide the analysis of the UV observations. 

\subsubsection{Line profile dependence on physical parameters}\label{analine}

Models ``$a$'', ``$b$'', ``$c$'' and ``$d$'' have been chosen in order to investigate the influence of the accretion rate on the line profiles. These models have the same white dwarf mass, but different accretions rates. Due to the dependence of the mass-loss rate on the effective temperature through Equation \eqref{eqeuler}, these models have different \.{M}$_w$. Consequently we have two effects, the change in wind density  and the change in wind and disk temperature, both arising from changes in the accretion rate (Eq. \ref{eqtemp}).

\begin{figure}[!b]
\centering
\includegraphics[height=0.5\columnwidth,width=0.5\columnwidth]{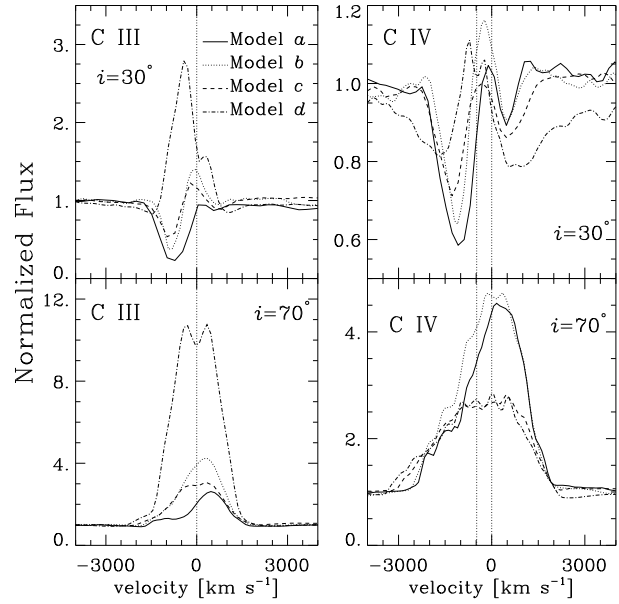}
\caption{Effect of accretion rate on the line profiles for \ionciii\ and \ionciv. \.{M}$_a$ = 1.0$\times$10$^{-8}$ M$_\odot$ yr$^{-1}$ (solid line), 5.0$\times$10$^{-9}$ M$_\odot$ yr$^{-1}$ (dotted line), 1.0$\times$10$^{-9}$ M$_\odot$ yr$^{-1}$ (dashed line) and 5.0$\times$10$^{-10}$ M$_\odot$ yr$^{-1}$ (dot-dashed line). The orbital inclination for all models in the upper panels is $i$=30$^\circ$ and $i$=70$^\circ$ in the lower panels.}
\label{fglines1}
\end{figure}

In figure~\ref{fglines1}, the variation in line profiles with accretion rate is shown for \ionciii\ and \ionciv\ UV lines. The accretion rate acts in a different way on each line but generally the blue absorption component is deeper when the accretion rate is higher. When the accretion rate is high the mass-loss rate is also higher and there are more absorbers in the wind. Further, an increase in the accretion rate increases the disk temperature and hence the wind ionization, and the relative intensities between lines changes accordingly. For instance, the intensity of a low ionization line like \ionciii\ increases when \.{M}$_a$ decreases, whereas the intensity of \ionciv\ decreases. This effect is clearer when the same comparison is made for the high orbital inclination case (lower panel in Figure \ref{fglines1}). We conclude that for high ionization lines \ionciv\ and \ionnv\, the stronger the accretion rate, the stronger the line intensity. For lower ionization energy lines (e.g., \ionskivb\ and \ionciii)  we found an increase in flux for lower mass accretion rates.

The effect of the primary mass is analyzed by comparing models ``$c$'' and ``$h$''. These disk models have, approximately, the same maximum disc temperature (Table \ref{tbmodels}). However, because of  dependence of wind parameters on  \mwd, the system with lower primary mass (0.6 \msun) has a  larger mass-loss rate ($\sim$ 6 times) than the system with \mwd\ =1 \msun. At a fixed temperature the WD mass influences both the disk gravity and size of the disk. Increasing the WD mass increases the gravity at a given ($r$/\rwd) disc location --- directly because of the increase in \mwd, and indirectly because \rwd\  has decreased. An increase in gravity translates into an increase in escape velocity, and hence a reduction in the mass-loss rate per unit area from a given location (as prescribed by $r/$\rwd) (see expression for $\dot{m}_o$, equation \ref{eqeuler2}). In addition, an increase in the WD mass decreases the size of the emitting wind region because of the dependence of $T_{eff}$ on ($r/\rwd$) (equation \ref{eqtemp}), which also leads to a reduction in the total mass loss from the disc. Similarly, the CAK model predicts that the wind terminal velocity will increase with  \mwd, as seen in Table~\ref{tbmodels}.

\begin{figure}[!t]
\centering
\begin{tabular}{cc}
\includegraphics[height=0.2\linewidth,width=0.23\linewidth]{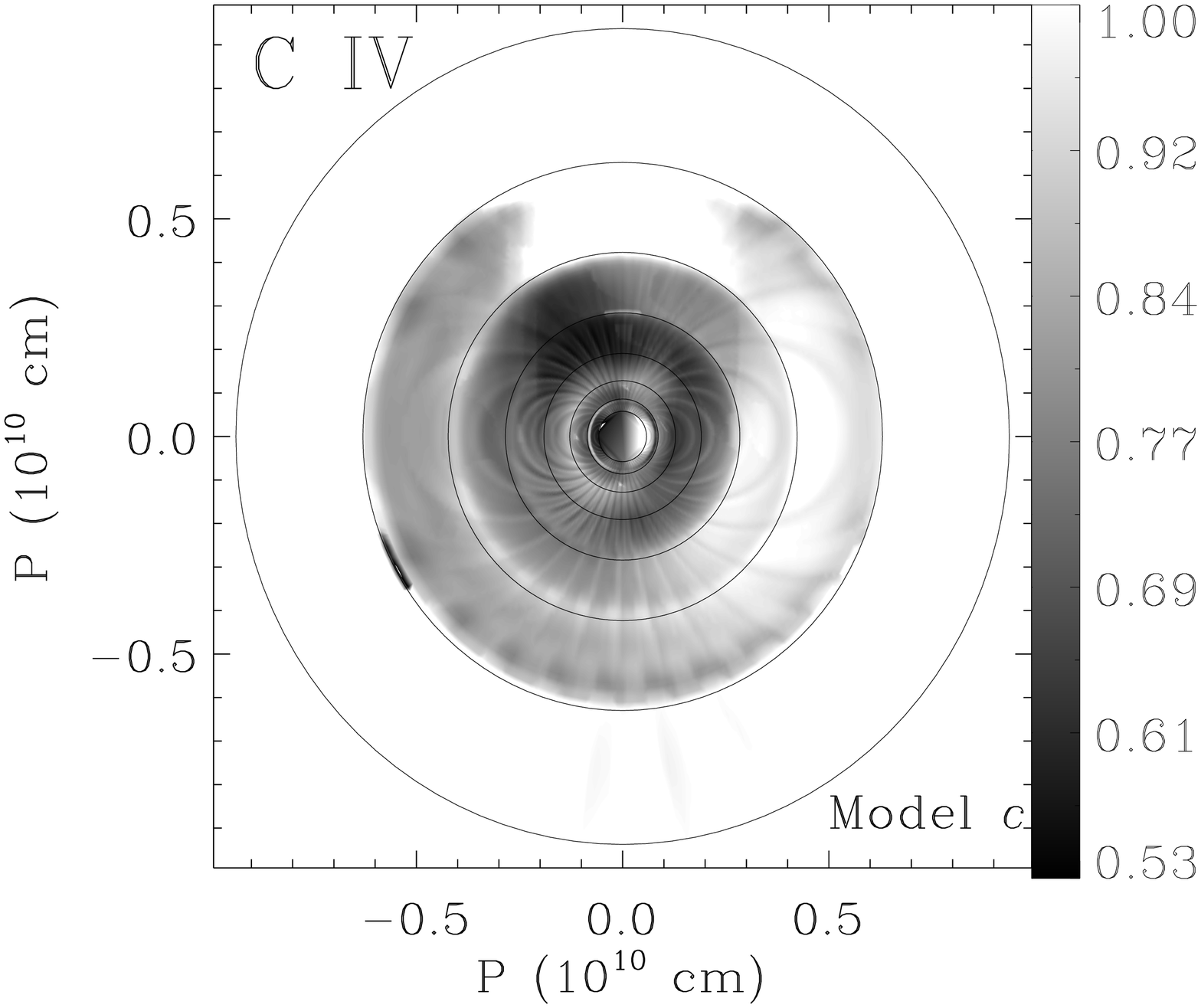} & 
\multirow{2}{*}[2.8cm]{\includegraphics[height=0.45\linewidth,width=0.25\linewidth]{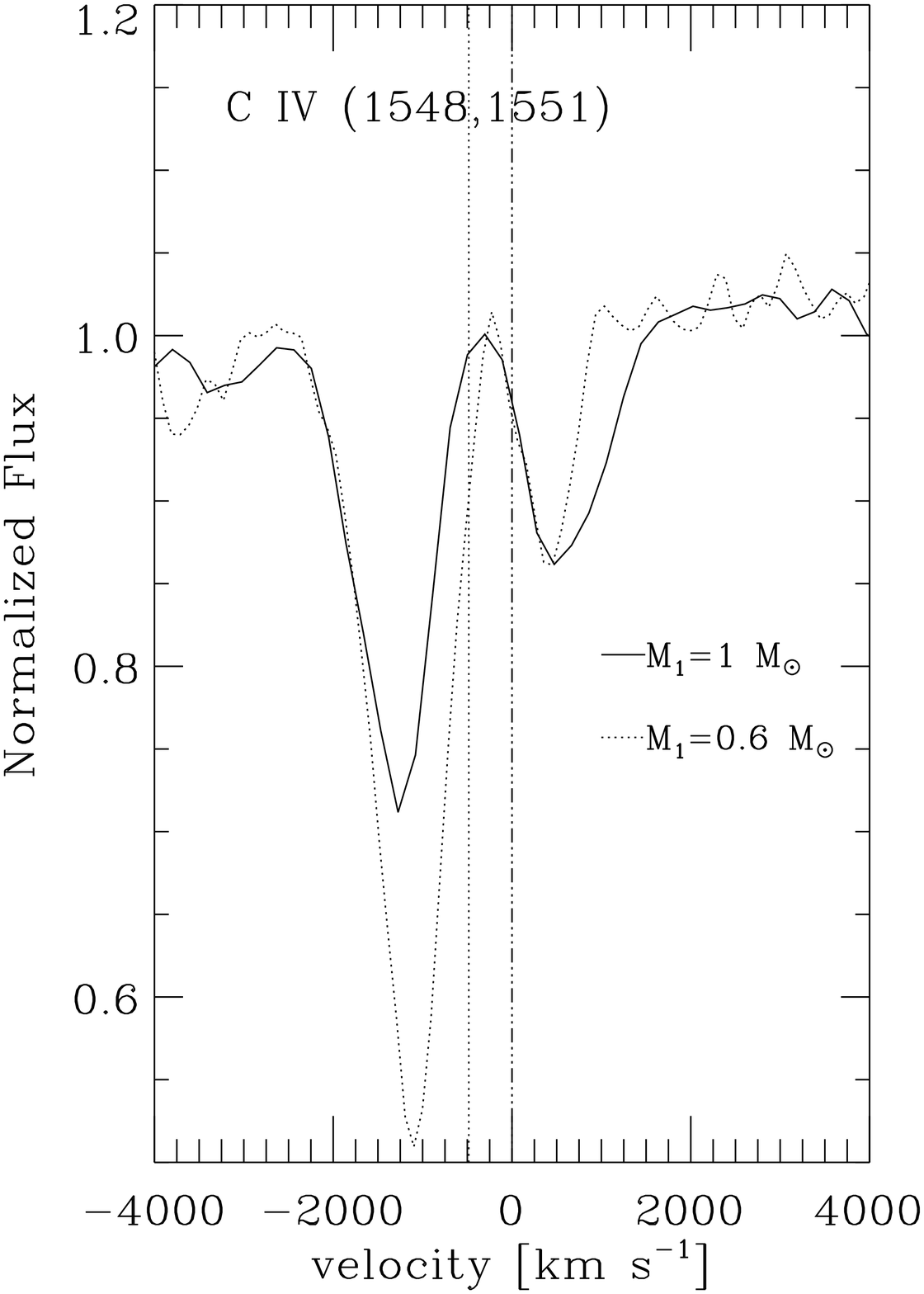}} \\
\includegraphics[height=0.2\linewidth,width=0.23\linewidth]{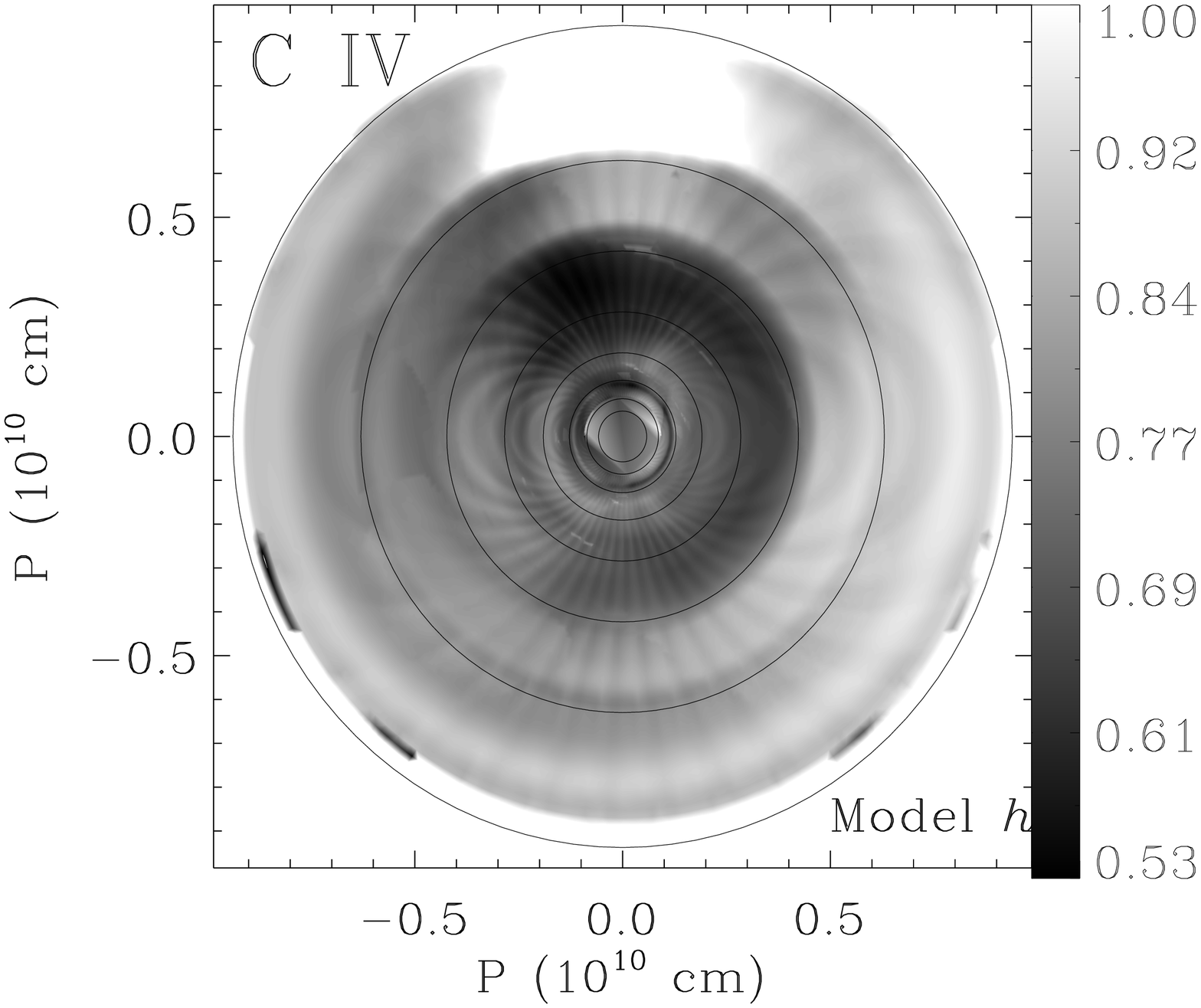} 
\end{tabular}
\caption{The ratio of the mean specific intensities in the line and in the continuum in the spectral region of  \ionciv\ for models ``$c$'' and ``$h$'' is shown for an observer's inclination of 30$^\circ$. Only the absorption regions ($I_l/I_c<1$) are shown. The right panel shows the calculated line profiles obtained by integrating across the whole disk. The models have similar disk temperatures but different primary masses, 1 \msun\ (solid line) and 0.6 \msun (dotted line).}\label{fgzone}
\end{figure}

The effect is evident in Figure~\ref{fgzone}, where the ratio of mean specific intensities in the line absorption (I$_l$) component over the continuum (I$_c$),  defined as

\begin{equation}\label{eqintcont}
\frac{I_{l}}{I_{c}}=\frac{\left[ \frac{1}{(\nu_{f_l}-\nu_{o_l})}\int^{\nu_{f_l}}_{\nu_{o_l}} I_\nu d\nu \right]}{\left[ \frac{1}{(\nu_{f_c}-\nu_{o_c})}\int^{\nu_{f_c}}_{\nu_{o_c}} I_\nu d\nu \right]}  \; \text{,}
\end{equation}

\noindent
is shown in the impact parameter space. In Equation \eqref{eqintcont}, the mean specific intensities are calculated between two frequencies that embrace the blueshifted absorption profile ($\nu_{o_l}$, $\nu_{f_l}$), and two frequencies that embrace the nearby continuum ($\nu_{o_c}$, $\nu_{f_c}$). Note that the absorption region (I$_l$/I$_c$$<$1) is larger when the primary mass is lower, and consequently, the absorption profile is deeper as it is shown in the right panel. The same behavior is found for the other UV resonance lines.  The intensity of the emission component depends mainly on the ionization state. For high ionization species, like \ionn{C}{iv} and \ionn{N}{v}, the intensity does not correlate with the primary mass. For the low ionization species and \ionskivb\ the emission component appears stronger by increasing \mwd. These lines are formed in the cooler external wind regions that are strongly influenced by the outermost wind radius $r_f$ and the external aperture angle $\theta_2$.

\begin{figure}[!b]
\centering
\includegraphics[height=.5\linewidth,width=.5\linewidth]{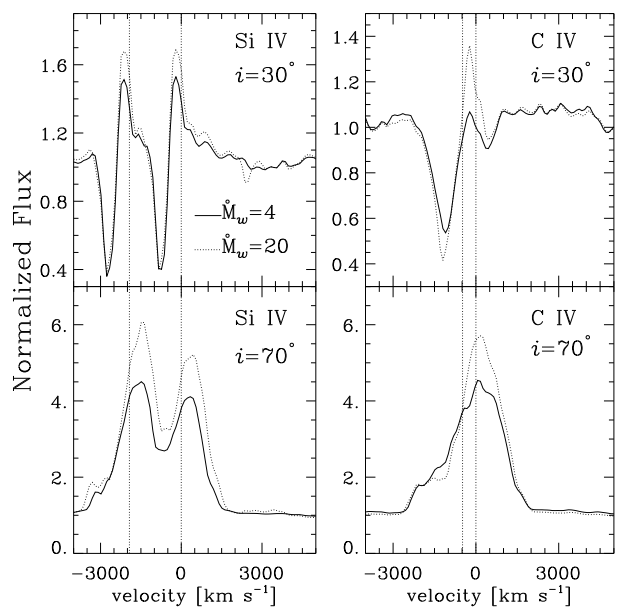}
\caption{Effects of the wind mass-loss rate on the line profiles of \ionskivb\ and \ionciv. Synthetic profiles are shown for model ``$h$'' with \.{M}$_w$=4 (solid line) and \.{M}$_w$=20 (dotted line), both in \massrate{1}{-11} units. Both models were calculated with an orbital inclination of  $i$=30$^\circ$ (upper panel) and $i$=70$^\circ$ (lower panel).}\label{fgmloss}
\end{figure}

We introduce a direct change in the local \.{M}$_w$ value aiming to study the effects that it could produce on the line profiles. This could be used to understand a more complex wind structure. The  physical and geometrical parameters of model ``$h$'' were used with a wind mass-loss rate increased five times (to \.{M}$_w$=\massrate{2}{-10}) for the comparison model. Figure \ref{fgmloss} shows the effects of that density enhancement on line profiles for two inclinations. The main effect is a rise in the line emission. For $i$=30$^\circ$ there is  also a  slight increase in the depth of the absorption profiles --- the profiles shown for the $i$=70$^\circ$ do not show absorption components.. These effects are expected from the increased wind density. The temperature and ionization structures are not much affected by the enhanced  mass loss rate.  Those lines formed in the wind from inner disk regions are more affected by variations in \.{M}$_a$ as these regions have stronger winds. In our models the extension of the line wings is unaffected by the mass loss.

\begin{figure*}[!t]
\centering
\includegraphics[height=0.4\linewidth,width=0.45\linewidth]{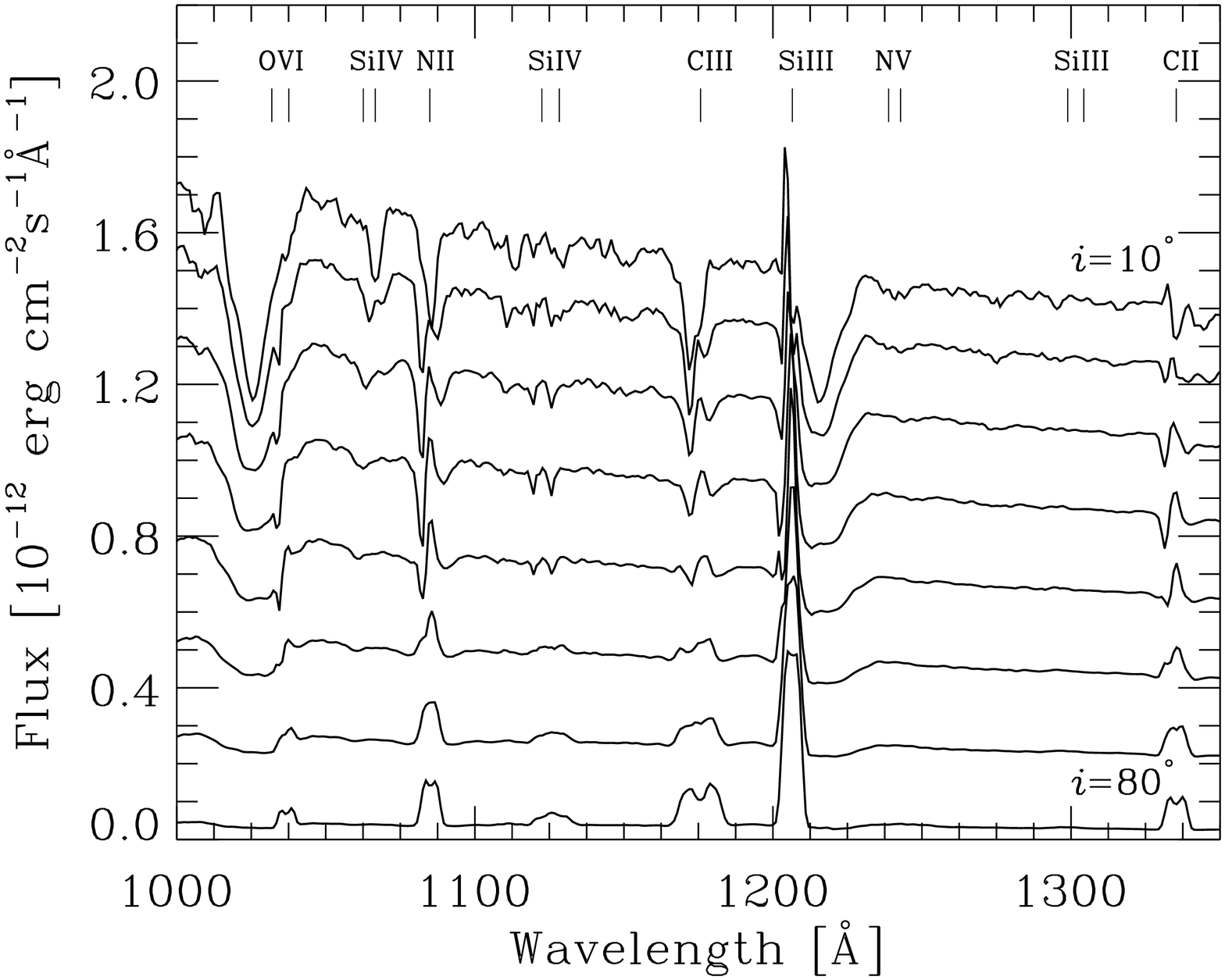}
\includegraphics[height=0.4\linewidth,width=0.45\linewidth]{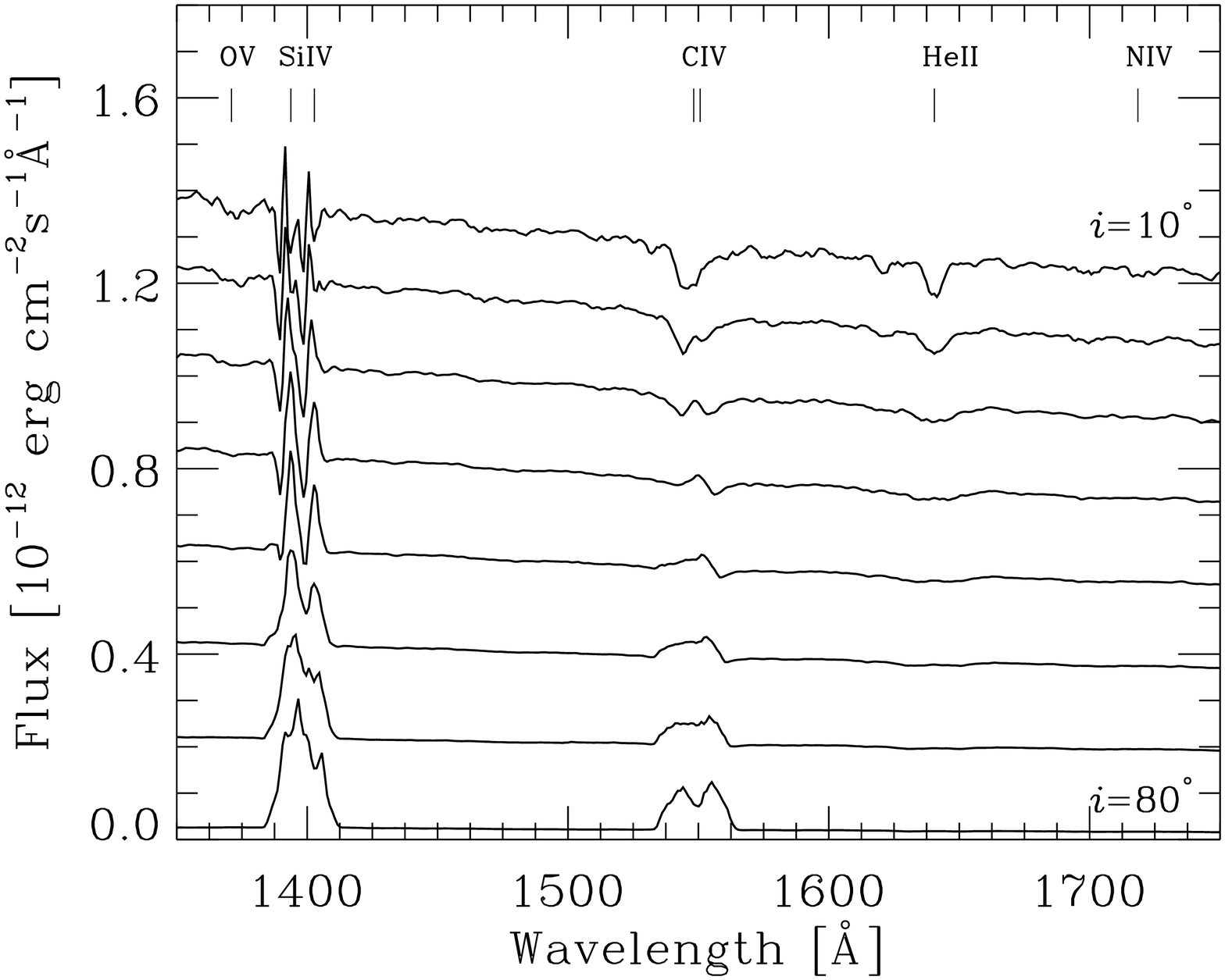}
\caption{Mid and far ultraviolet disk model spectra at several orbital inclination angles for model ``$d$''. The angles span the values from 10$^\circ$ to 80$^{\circ}$, with 10 degree steps. The spectra are scaled to a distance of 400 pc and separated by a constant offset of 1.5$\times$10$^{-13}$ erg cm$^{-2}$ s${-1}$ \AA$^{-1}$ for clarity.}\label{fgincl}
\end{figure*}

The effect of the orbital inclination on the synthetic spectra was extensively explored. In  Figure~\ref{fgincl} the synthetic spectra are shown for model ``$d$'' with several inclinations. The inclination angle goes from 10$^\circ$ to 80$^\circ$. The overall behavior follows the observational data for CVs in the UV. Low inclination models present line profiles that are preferentially in absorption. The atmospheres are seen almost face on, the profiles are blueshifted due to the strong vertical acceleration, and the line profiles are narrow because the influence of the wind rotation is weak. In the case of intermediate inclinations, emission structures begin to emerge in the profiles because high temperature wind regions are now frontwards of the cooler disk continuum. These emission structures produce P~Cygni profiles in several lines such as \ionciii, \ionskivb\ and \ionnii. The emission structures appear blueshifted in intermediate inclination models because the wind regions with positive projected velocities are occulted by the disk, especially in the case of high ionization lines that are produced in the inner disk regions. When the orbital inclination increases, the continuum source aspect decreases relative to the wind dominated region. In this sense, for high orbital inclination models the continuum dims as the emission line intensity grows as shown in Figure~\ref{fgincl}. The line width is also increasingly dominated by disk-wind rotation in high inclination models, where the emission lines become broader and the blending with photospheric profiles is stronger. The UV continuum intensities strongly depend on disk inclination via the limb darkening effect \citep{diaz96}. When the mass-loss rate is strongly increased a higher reddening is found, but this effect is weaker in the case of high orbital inclinations.

All models show emission in \ionskiiia\ even in low inclination disks. Also, emission components were found in low inclination models for \ionskivb. These emission components are not found in most low inclination CVs. In our models these features are mostly produced in the outermost wind regions, from gas that is coming from cooler portions of the disk ($r$$\sim$$r_f$). The wind electron temperature there is $\sim$11700 K, and the dominant ion of silicon is \ionn{Si}{iii} but with a fraction of \ionn{Si}{iv}. The outer disk atmospheres are larger in these models, and that is the reason why the emission is present even for low inclinations. A similar situation happens for \ionciii, but in this case the
\ionn{C}{iii} fraction in the outermost atmosphere is lower and the corresponding emission structures are weaker.

The general behavior of the observed line profiles with the orbital inclination is well reproduced by models. However, these models yield some profile structures that are not commonly found in the UV lines of CVs, like strong \ionskivb\ emissions and a blueshifted emission for \ionciv\ and \ionnv\ in low inclination systems. These features were found to be dependent on the wind geometry and ionization structure that bears different emission volumes for each ionization state.

Aiming to study the influence of an extended low temperature wind filling part of the primary Roche lobe, extra models were calculated with the atmosphere extending beyond the outermost region of the original model. This additional atmosphere wind model has a lower temperature and its structure is also extended until R$_{disk}$ and limited by $\theta_2$ in the way explained in \S \ref{method}. The presence of this atmosphere shows a significant effect on the synthetic spectrum. The major changes occur in the low ionization lines (\ionciii, \ionskiiia\ and \ioncii) and also in \ionskivb. These lines have their emission structures weakened while their absorption profiles become deeper. The high ionization lines (eg. \ionciv) are not affected by the inclusion of the cooler wind models. This simulation reveals the importance of the outer wind ionization structure, and helps us to constrain the parameters that characterize the wind and disk.

\section{COMPARISON WITH OBSERVATIONS}\label{observations}

In this section we confront our synthesized spectra with HST UV observations of the Nova-like CVs RW Tri and the IUE data for V347 Pup. Through this comparison, the capabilities and limitations of our method are revealed. The need for some model changes became evident from this comparison, and consequently some additional calculations were made. The new models were labeled as ``$b'$'',``$h'$'', ``$b''$'' and ``$h''$''. The model parameters are shown in table \ref{tbmodpar}.

\subsection{RW Tri}\label{rwtri}

The Nova-like RW Tri is an eclipsing system. The orbital period is $\sim$0.25 days ($\sim$5.5 hours), the primary mass is around $\sim$0.55 \msun\ and the orbital inclination is $\sim$ 70$^\circ$ \citep{ritter03,poole03}. The distance was estimated by \cite{mcarthur99} to be between 310 and 380 pc. From a UV continuum analysis, \cite{puebla07} calculated a value for the mass accretion rate of $\sim$\massrate{4.6}{-9}. This value is compatible with those calculated by \cite{groot04} and \cite{mizusawa10}. A summary of binary system parameters is shown in Table~\ref{tbsyspar}. The UV as well as the optical spectrum show strong emission lines. Observations in the UV also show that the lines are less eclipsed than the continuum \citep{cordova85,drew85}. 

The spectroscopic data was obtained from the HST data archive (MAST\footnote[1]{http://archive.stsci.edu}). All spectra were taken by the GHRS spectrograph in RAPID mode,  using the grid G140L with a spectral coverage between 1150 \AA\ and 1660 \AA\ in low resolution ($\sim$1 \AA) \citep[for more details of data see][]{mason97}. The data were corrected for interstellar extinction using the \cite{cardelli89} law and a value of $E(B-V)$=0.1 \citep{bruch94}.

In this work we used the basic parameters of the ``$b$'' and ``$h$'' models from Table \ref{tbmodels} in order to fit the RW Tri UV data. The ``$h$'' model has physical parameters close to those previously estimated for RW Tri. Unlike, model ``$b$'' has a white dwarf mass higher than reported for RW Tri. We have also analyzed this case because this high mass model is capable to produce stronger \ionn{C}{iv} and \ionn{N}{v} emission. It was necessary to change some other parameters in order to fit the emission lines. In particular, the geometrical parameter $\theta_2$ was adjusted in order to match the observed intensity of \ionciii\ and \ionskivb\ lines. The $\theta_1$ value is used for matching the \ionnv\ and \ionciv\ lines. Also, in order to achieve better line ratios, we have arbitrarily included denser wind regions as explained in section \ref{analine}, by enhancing the local mass loss rate five times between certain radii ($r_{d_1}$ and $r_{d_2}$).

\begin{figure*}[!t]
\centering
\includegraphics[height=0.4\linewidth,width=0.8\linewidth]{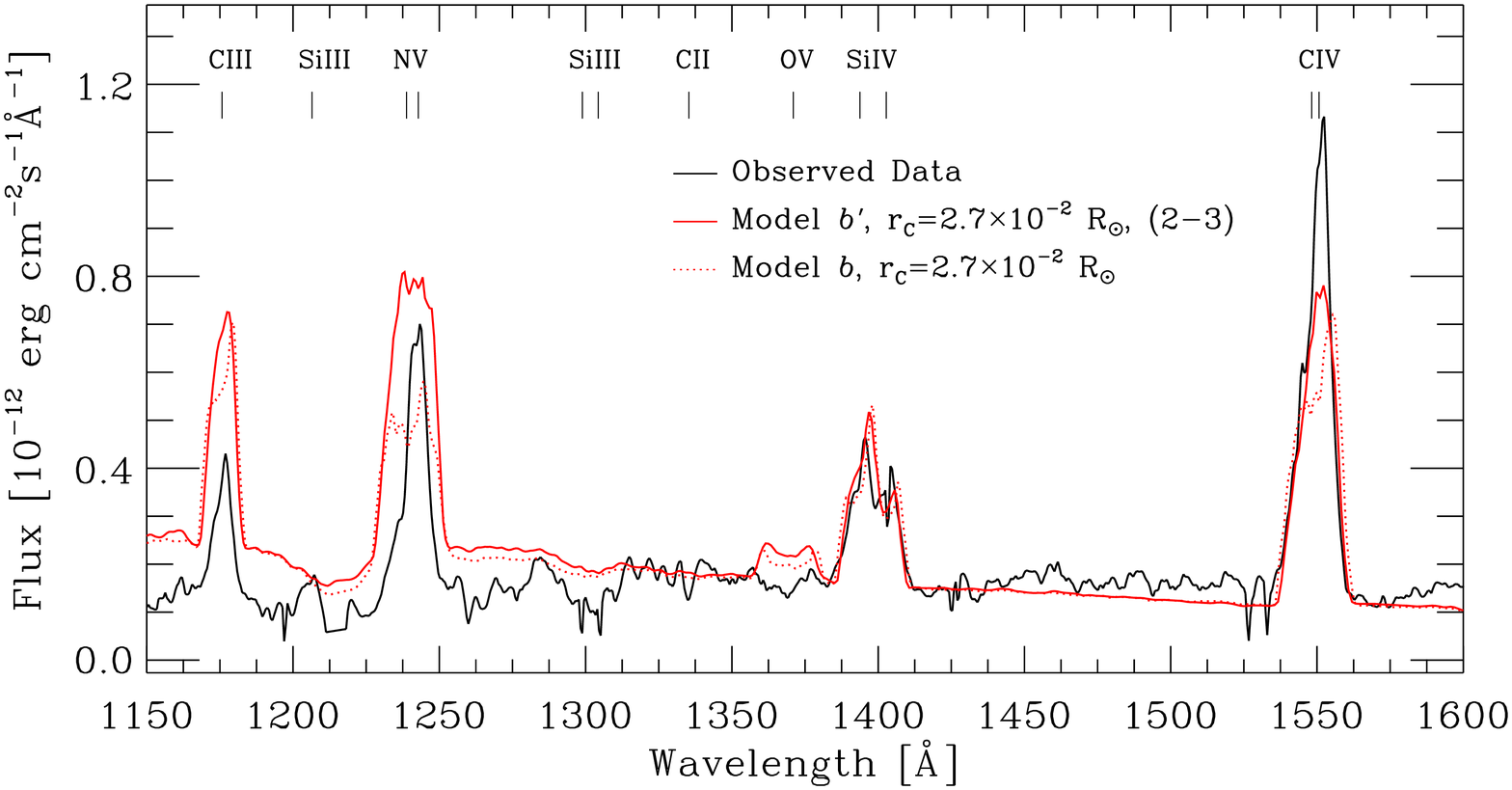}
\includegraphics[height=0.4\linewidth,width=0.8\linewidth]{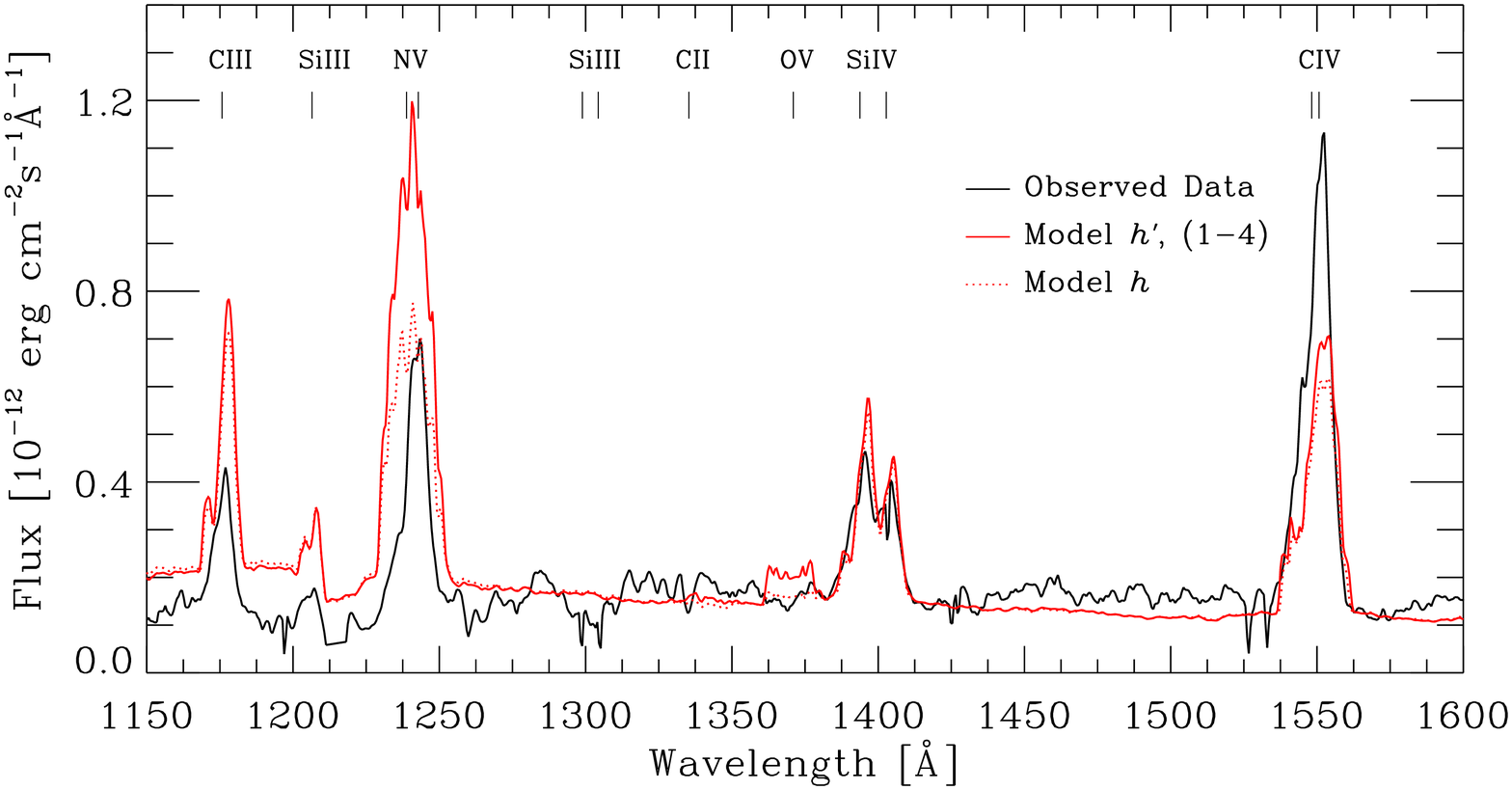}
\caption{Best simulations for RW Tri UV data. The data are the mean of spectra taken out of eclipse. The upper panel shows the best-fit models using the disk parameters for model ``$b$'' and ``$b'$'' (see text for details). The lower panel shows the best-fit models using the disk parameters for model ``$h$'' and ``$h'$''. The numbers in the parentheses show the rings where a density enhancement was included for the altered models (see text for details).}\label{fgrwtri1}
\end{figure*}

In the top panel of Figure \ref{fgrwtri1} the UV data for RW Tri is shown together with the ``best'' calculated synthetic spectra using the ``$b$'' model and the modified model ``$b'$''. Synthetic spectra were computed assuming an orbital inclination of $i$=70$^\circ$ and inner aperture angle $\theta_1$=5$^\circ$ and $\theta_2$=35$^\circ$. The outermost radius for the wind atmospheric structure is $r_f$=17 \rwd\ ($\sim$\sci{9.05}{9} cm) with $T_{eff}(r_f)$$\sim$16800 K. Both models in Figure \ref{fgrwtri1}a show the synthetic spectra for a model with $r_C$=\sci{1.87}{9} cm. The models were scaled to a distance of 550 pc. We set $r_C$ lower than $r_f$ because if $r_C$ is increased to $r_f$ ($\sim$\sci{9.05}{9} cm) the continuum flux is slightly increased, damaging the line-continuum ratio. The derived distance seems too high when compared with the distance estimates found in the literature ($\sim$350 pc). This is possibly due to the \mwd\ used in the models which is too high considering the range found in the literature. The modified model ``$b'$'' has a density enhancement between 1.56 \rwd\ and 2.32 \rwd\ (0.83-1.25 $\times$10$^9$ cm) set by increasing $\dot{m}(r)$ to 5 times the value given by our wind acceleration prescription. The figure shows the improvement on line emission ratios from model ``$b$'' (without dense region) to model ``$b'$'' (with dense region). The intensities of the \ionciv\ and \ionnv\ have been enhanced, as expected, but still are weaker than those observed. A significant improvement was found for the \ionskivb\ line, that shows a better line ratio regarding \ionn{C}{iv} and \ionn{N}{v} in model ``$b'$''.

The lower panel of Figure \ref{fgrwtri1} shows the best spectra using the model ``$h$'' and the modified model ``$h'$''. The inner and outer aperture angles are $\theta_1$=2$^\circ$ and $\theta_2$=10$^\circ$. This model has a \mwd\ value closer to that estimated for RW Tri yielding a scaled distance of $\sim$450 pc, closer to the literature values. In the ``$h'$'' model a dense region was again included to improve the lines as shown in Figure~\ref{fgrwtri1}, in the same sense as explained for the ``$b'$'' model. In the figure, the values between parentheses are the positions (in ring numbers) of the density enhanced region (table \ref{tbmodpar}). The models $h's$ are cooler than $b's$ models so their continua are redder, but still bluer than the observed continuum. We also found an improvement in the \ionciv\ and \ionnv\ line intensities in $h'$ model, but for the case of \ionn{C}{iv}, it was not enough to match the observations.

\begin{figure}
\centering
\includegraphics[height=0.5\columnwidth,width=0.5\columnwidth]{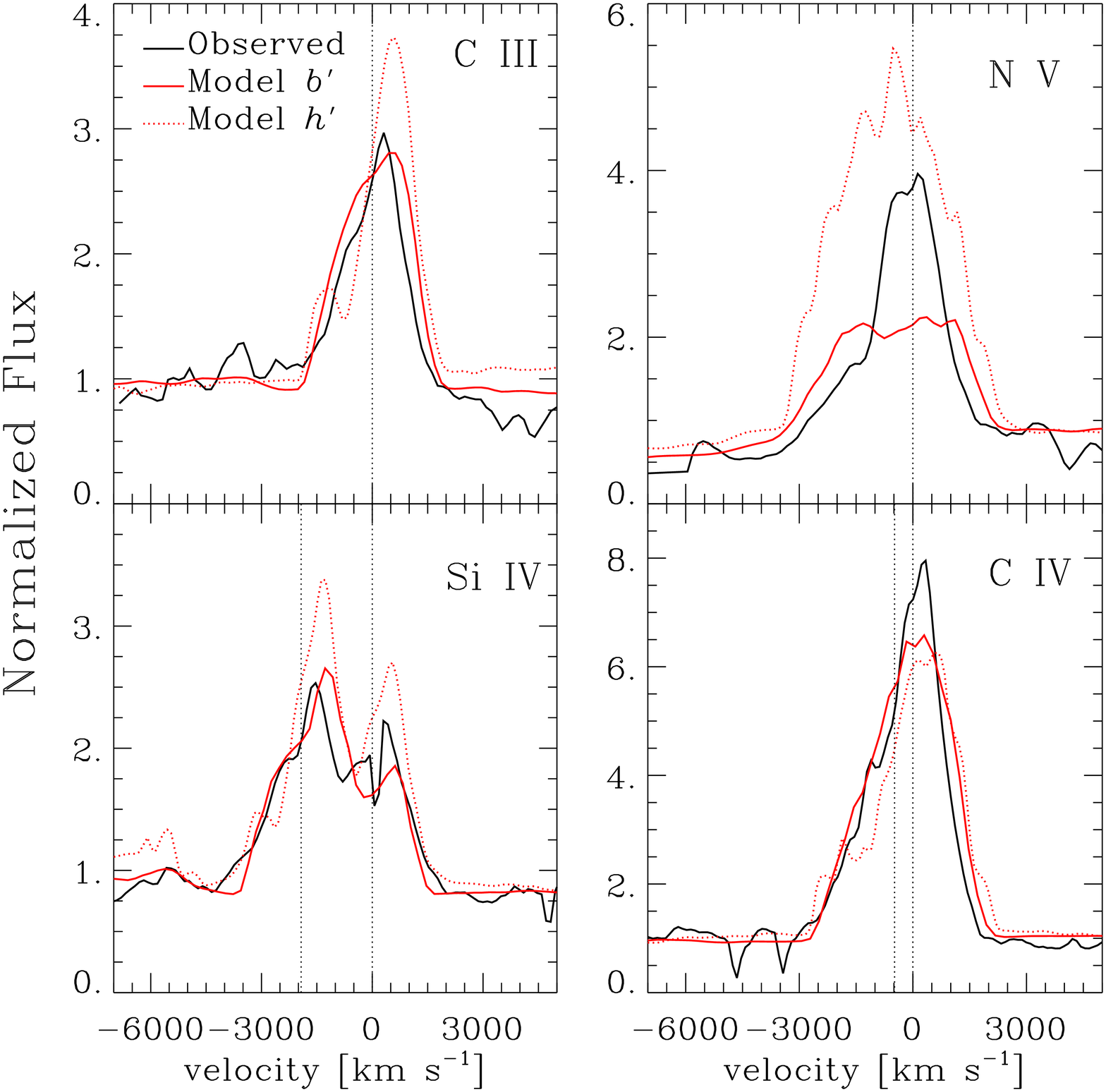} 
\caption{Continuum normalized line profiles for the main UV lines of RW Tri: \ionciii, \ionnv, \ionskivb\ and \ionciv. The parameters for each model are given in the text.}\label{fgrwtri2}
\end{figure}

In Figure \ref{fgrwtri2} the emission profiles for the main UV lines are shown together with the better synthetic profiles (models $b'$ and $h'$). We found a good agreement for the \ionciii\ and \ionskivb\ profiles. The model with higher \mwd\ displays better agreement with the line widths. Both models show a lack of flux in the \ionciv\ line, even when the inner wind regions are enhanced. This indicates that the \ionn{C}{iv} emission region is larger than predicted by our models. We conclude that the boundary layer emission may play an important role in the formation of such a strong \ionn{C}{iv} line. Model \ionnv\ lines are wider than the observed profiles showing that the \ionn{N}{v} line in the system is possibly coming from a region with a lower rotation velocity. The model ``$b'$'' line intensity is lower, as in the case of \ionn{C}{iv}, but it is stronger in model $h'$ which has a more extended (in radius) dense region (see table \ref{tbmodpar}). It is clear from this that the high ionization region in the wind should be larger. The models with \mwd\ =0.6 \msun\ show a narrow blueshifted emission structure for \ionciii\ and \ionskivb\ lines. This component comes from the innermost regions with high vertical acceleration.

Table \ref{tbrwtri} shows the equivalent widths (EWs) for model ``$b'$'', ``$h'$'' and data. An evaluation of the model versus observed profiles is also given, on the basis of the integrated residuals and visual inspection. We can see that model ``$b'$'' shows better values of EWs and also a better profile matching for \ionciii\ and \ionskivb. On the other hand, the high ionization lines \ionnv\ and \ionciv\ are not so good. Recently, \cite{noebauer10} modeled the UV lines of RW Tri using an improved version of the \cite{long02} code developed by  \cite{sim05}. Their models incorporate the formation of recombination lines \citep[see also][]{lucy02,lucy03}. They found good agreement for \ionn{C}{iv} and \ionn{Si}{iv} line intensities, but with a mass-loss rate almost 5 times larger than ours. Although this difference their and our values are compatible with the hydrodynamical models predictions. Their 3D Monte-Carlo model has a reduced \ionn{Si}{iv} emission volume, yielding a better match of the \ionn{Si}{iv} and \ionn{C}{iv} flux ratio. They also obtain a strong \ionn{C}{iv} emission without include dense regions. In their models \ionn{C}{iv} fills an extended region in the wind, \ionn{N}{v} is limited to inner wind regions, \ionn{Si}{iv} and \ionn{C}{iii} and lower ionization species are limited to lower wind regions (close to disk surface) at larger radii (see fig 8 in \cite{noebauer10}). 

\begin{figure}
\centering
\includegraphics[height=0.8\columnwidth,width=0.8\columnwidth]{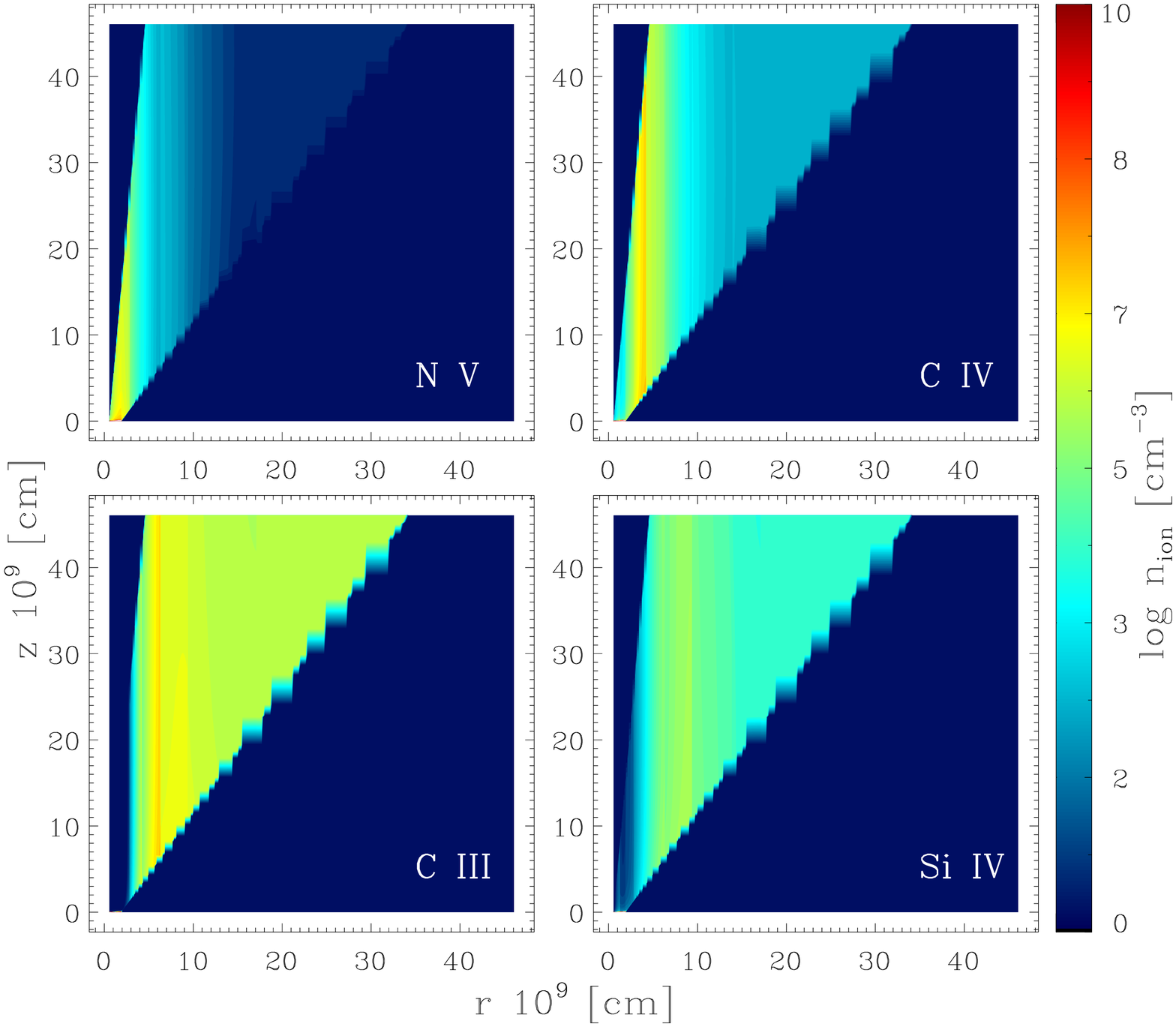} 
\caption{Ion density structure for model ``$b'$'' with r$_C$ = \sci{2.7}{-2} \rsun. The number density in cm$^{-3}$ is shown for \ionn{N}{v}, \ionn{C}{iv}, \ionn{C}{iii} and \ionn{Si}{iv}. }\label{fgrwtri3}
\end{figure}

In figure \ref{fgrwtri3} we present the ion density structure calculated with using the model ``$b'$'' parameters. The strong dependence of the ionization structure on the temperature of radiation that is coming from disk is evident. In our approximate geometry this radiation arises only from disk regions that are strictly below of each wind point. That is the reason why the ionization structure shows larger variations in the radial direction than those found in the vertical profiles. In our model structure there is no radial radiation transfer. Distinctly, \cite{noebauer10} with their Monte Carlo method obtained a 3D treatment of the extended wind regions assuming a flat disk surface. These authors found a more extended \ionn{C}{iv} emitting region thanks to the irradiation of the distant wind by the inner disk.

\subsection{V347 Pup}\label{v347pup}

The luminous CV V347 Pup is a high orbital inclination \citep[$i$$\sim$80$^\circ$][]{ritter03} Nova-like system, with bright optical and UV emission lines. Its orbital period is $\sim$0.23 days ($\sim$5.5 hours), the primary mass is $\sim$0.63 \msun\ and the distance is estimated as $\sim$510 pc \citep{diaz99,thoroughgood05}. \cite{puebla07} estimated a mass accretion rate of $\sim$\massrate{6}{-9}, a value compatible with that  calculated later by \cite{ballouz09}. A summary of the binary parameters is shown in Table \ref{tbsyspar}. The UV data were obtained from the IUE archive and were collected with the SWP camera in large aperture mode with a spectral resolution of $\sim$6 \AA. The spectra extend from 1150 \AA\ to 1950 \AA. The data were corrected for interstellar extinction using the \cite{cardelli89} law and a value of $E(B-V)$=0.06 \citep{mauche94}
 
\begin{figure*}[!t]
\centering
\includegraphics[height=0.4\linewidth,width=0.8\linewidth]{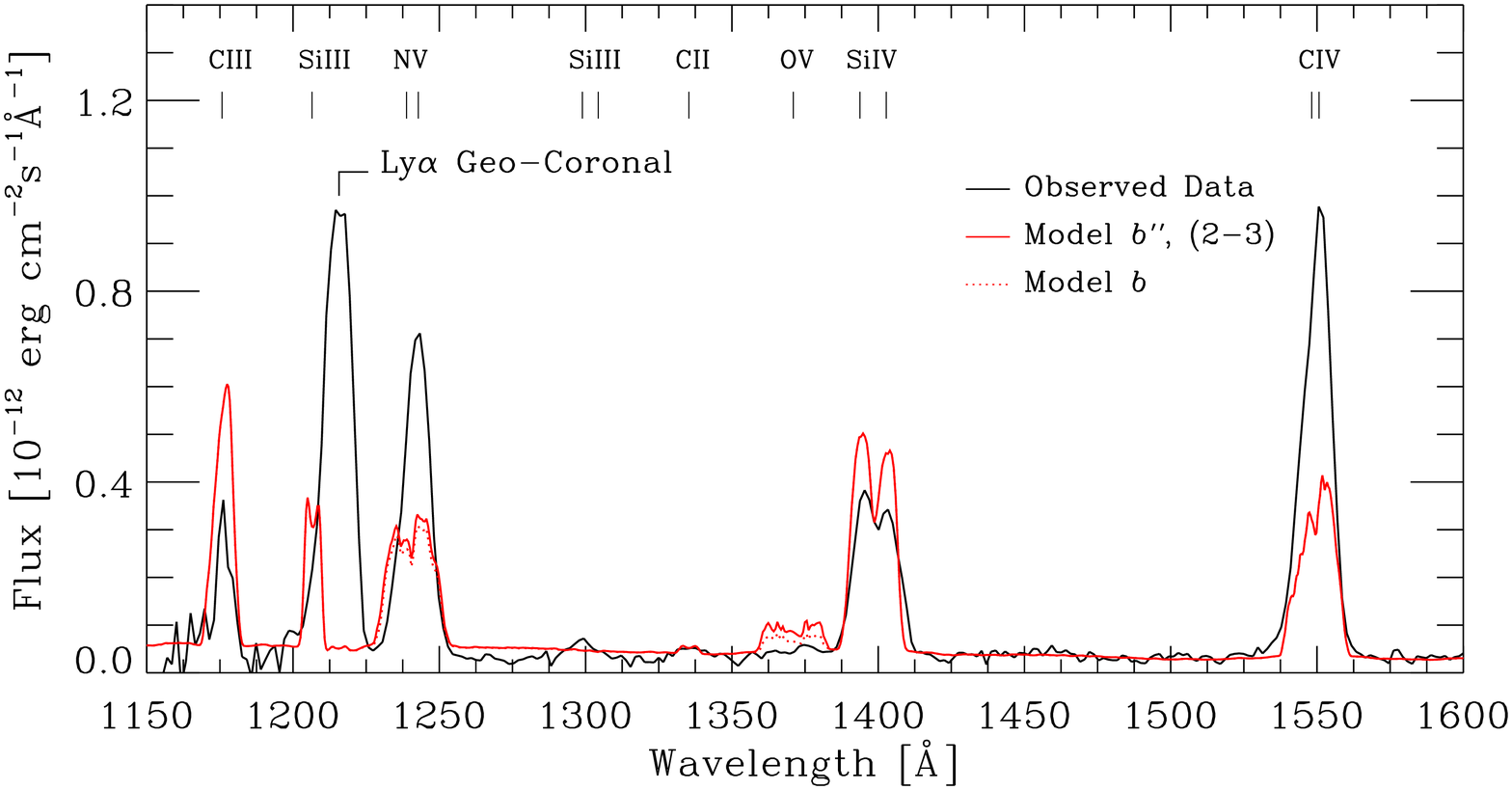}
\includegraphics[height=0.4\linewidth,width=0.8\linewidth]{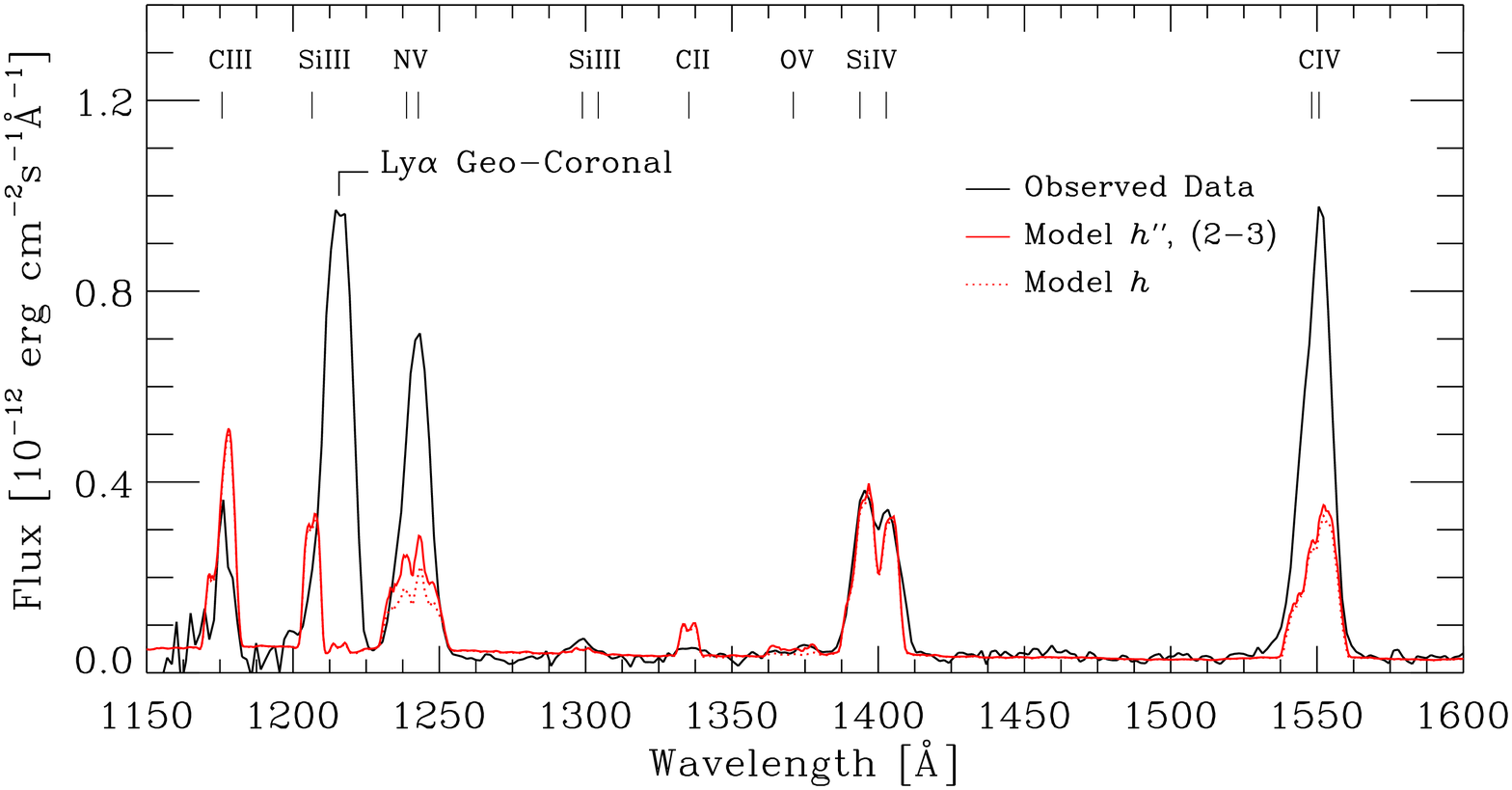}
\caption{Best simulations for V347 Pup UV data. The data are the mean of a spectra taken out of eclipse. The upper panel shows the best model using the disk parameters for model $b$ and $b''$. The lower panel shows the best-fit models using the disk parameters for model $h$ and $h''$. The numbers in the parentheses show the rings where a density enhancement was included (see text for details).}\label{fgv347pup1}
\end{figure*}

In this case we used the modified models ``$b''$'' and ``$h''$'' (see table \ref{tbmodpar}). Figure \ref{fgv347pup1} shows the comparison of model spectra with the V347 Pup IUE data. All models have the same geometrical parameters: $i$=80$^\circ$, $\theta_1$=5$^\circ$, $\theta_2$=45$^\circ$ (including  $b$ and $h$ models). As is the case for RW Tri, a dense region was included in order to improve the intensities of the high ionization lines. Following the previous models, this region is five times denser than the original model and is located between ring 2 and 3 (0.83-1.25$\times$10$^{-9}$ cm when \mwd\ =1 \msun\ and 1.25-1.73$\times$10$^{-9}$ cm when \mwd\ =0.6 \msun). However, for this case we did not find any relevant differences with the non-altered models ($b$ and $h$). We only found a slight increment in \ionnv\ and \ionciv\ lines.  The upper panel model in Figure \ref{fgv347pup1} was scaled to 710 pc to match the flux level data in the 1450 \AA\ continuum region. This model shows stronger lines than those observed for \ionciii\ and \ionskivb\ lines, and weaker lines for \ionnv\ and \ionciv. When using a primary mass of 0.6 \msun\ (lower panel in the figure) the model intensities of \ionn{Si}{iv} and \ionn{C}{iii} lines improve, while the intensities of \ionn{N}{v} and \ionn{C}{iv} lines decrease. This synthetic spectrum was scaled to a distance of 570 pc, a distance closer to that found in the literature, due to the lower \mwd.

\begin{figure}[!t]
\centering
\includegraphics[height=0.5\linewidth,width=0.5\linewidth]{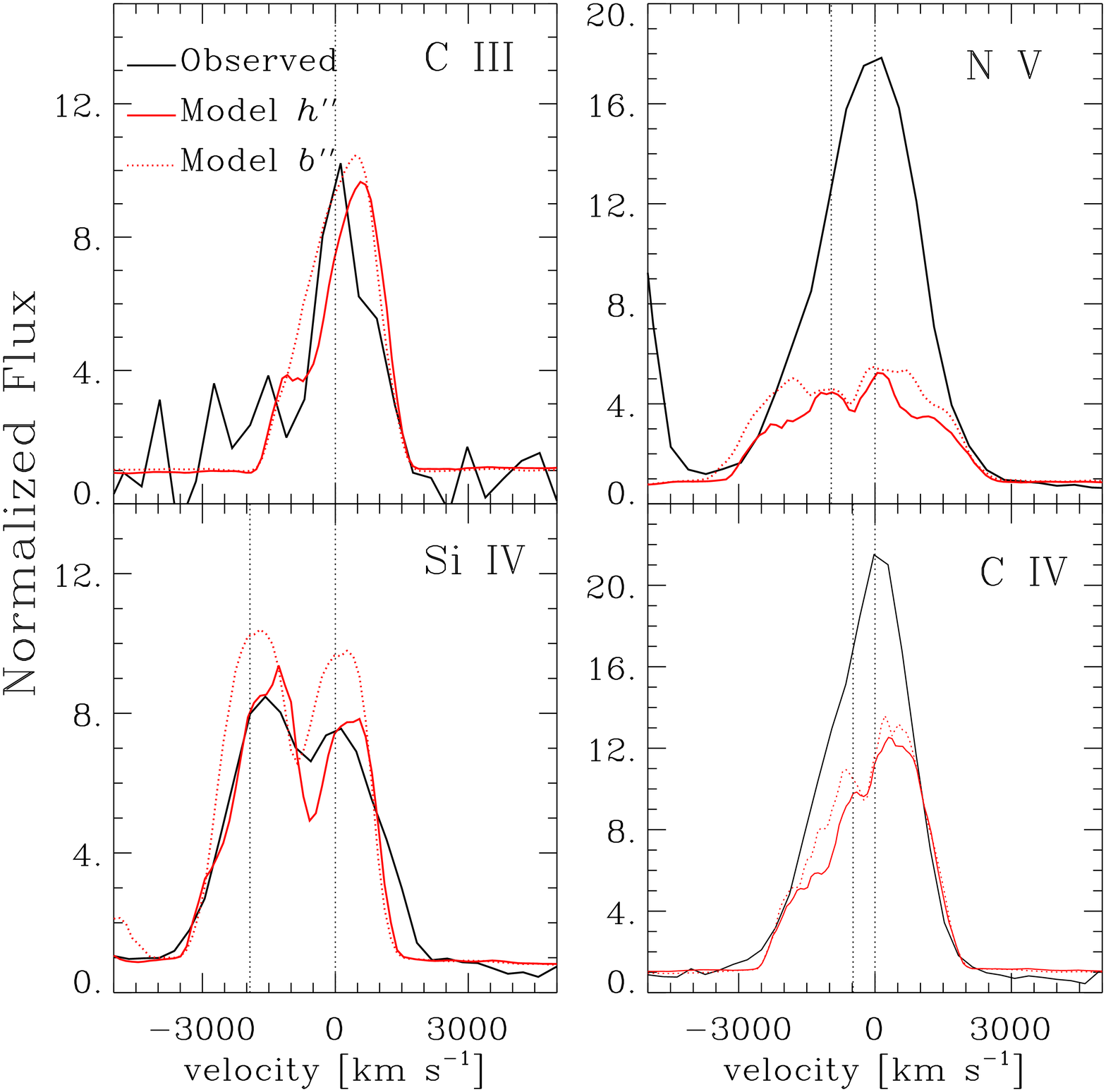}
\caption{Continuum normalized line profiles for the main UV lines of V347 Pup: \ionciii, \ionnv, \ionskivb\ and \ionciv. The parameters for each model are given in the text.}\label{fgv347pup2}
\end{figure}

Figure \ref{fgv347pup2} shows the line profiles of the models along with the data. A fair agreement for the \ionciii\ and \ionskivb\ profiles is found, but the models underestimate the flux in the high ionization lines (\ionnv\ and \ionciv). This model behavior can be explained by the temperature structure and by the emission volume. The density increment at the inner wind zones helps to enhance the high ionization lines emission, but not enough to match observations because of the limited volume where the high temperature radiation ionizes the ejected gas. The enhanced intensity in \ionn{C}{iii} and \ionn{Si}{iv} in the model with high \mwd\ is due to the larger emission volume of the outermost wind atmosphere, where these lines are mainly produced.

In Table \ref{tbv347pup} we compare EWs for  models ``$b''$'' and ``$h''$'' with observation. Similar to the RW Tri case, \ionciii\ and \ionskivb\ are well reproduced by the models while \ionnv\ and \ionciv\ are not. Again, we found a lack of flux for the last lines due to their small emission volume. \cite{shlosman96} modeled the \ionciv\ line profile in and out of eclipse. They could fit the line profiles, but needed a very high mass-loss rate of $\sim$\massrate{1}{-9}, almost half of the accretion rate, and used system parameters (\mwd\ and \.{M}$_a$) that produce a hotter disk. We modeled a cooler disk than that of \cite{shlosman96} (as they suggested) but could not attain the data flux line. It is possible that the inclusion of other ionization sources, such as the BL emission, could improve the match to the \ionnv\ and \ionciv\ emission profiles.

\begin{figure}
\centering
\includegraphics[height=0.8\columnwidth,width=0.8\columnwidth]{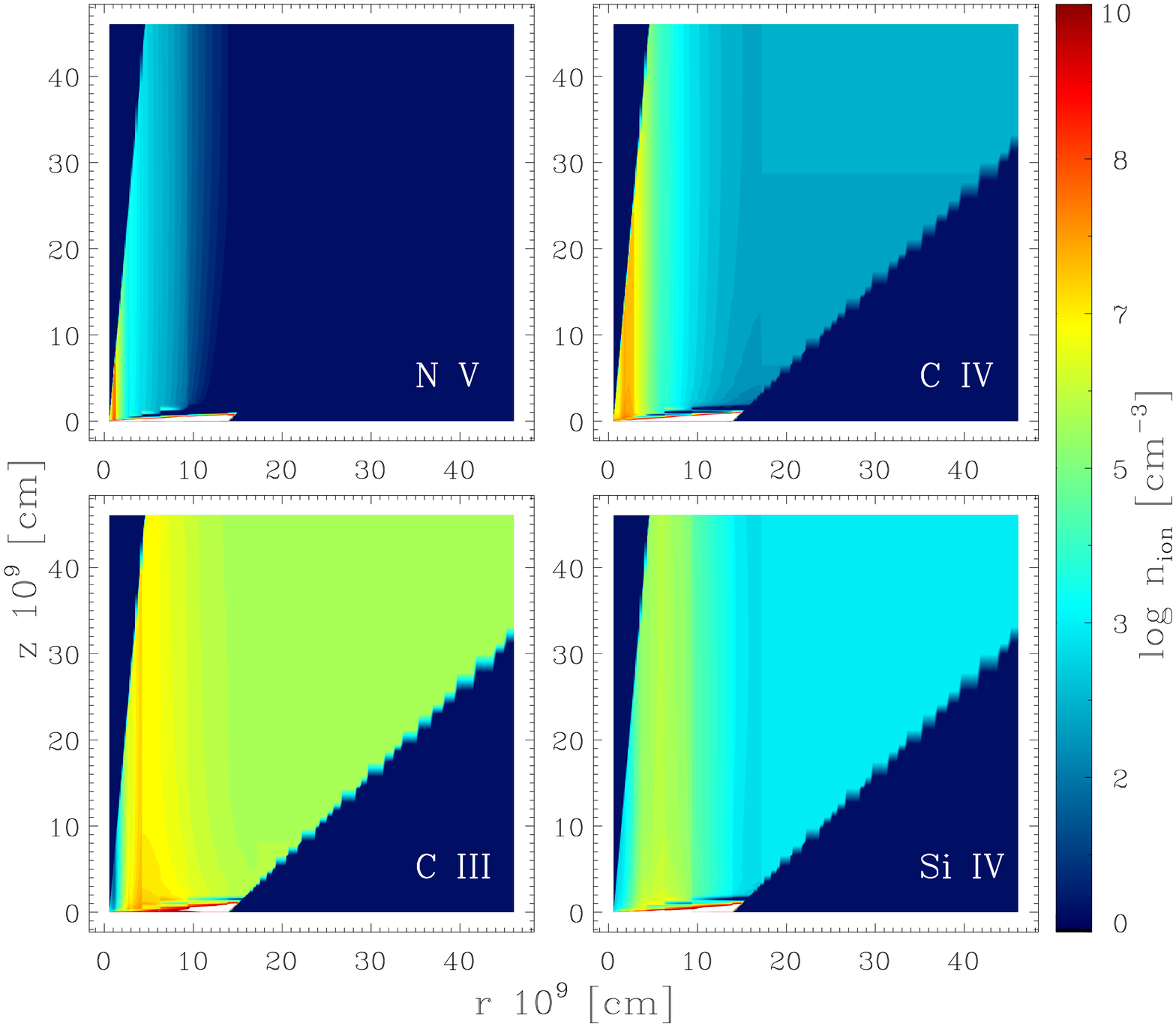} 
\caption{Ion density structure for model ``$h''$''. The number density in cm$^{-3}$ is shown for \ionn{N}{v}, \ionn{C}{iv}, \ionn{C}{iii} and \ionn{Si}{iv}.}\label{fgv347pup3}
\end{figure}

In figure \ref{fgv347pup3} we present the ion density structure calculated using the model ``$h''$'' parameters for \ionn{N}{v}, \ionn{C}{iv}, \ionn{C}{iii} and \ionn{Si}{iv}. In this case the ``$cylindrical$'' structure is also evident. In this cooler disk, the regions with dominant \ionn{N}{v} and \ionn{C}{iv} are closer to disk axis than they are in high white dwarf mass case case, resulting in weaker lines, unlike the \cite{noebauer10} models where, specially for \ionn{C}{IV}, these regions are fairly extended through the wind.

\section{DISCUSSION}\label{discussion}

In our simulations, the vertical wind velocity is calculated taking into account the local physical environment for each wind atmosphere. Previously, the disk and wind were treated independently or with a fiducial transition region. Also, in earlier works, the ionization equilibrium and level populations have been calculated using simplified methods. Even with several simplifying assumptions, we found that our model spectra resulting from our complete synthesis present emission lines and continuum fluxes close to those observed in bright CVs. By comparing observed data of two high inclination systems with the models we have found similarities between the model and actual structured line profiles. However, we include denser regions within the wind in order to obtain reasonable line intensities, in particular for matching the high ionization line fluxes. The presence of condensations was previously predicted by hydrodynamical models of accretion disk winds \citep{pereyra97,pereyra00,pereyra03}. The figures \ref{fgrwtri1} and \ref{fgv347pup1} show that it was not enough to exactly match the observed profiles. We found good agreement with observations for \ionn{Si}{iv} and \ionn{C}{iii} lines. For high ionization lines \ionciv\ and \ionnv\ the model lines are weaker than those observed. This could be due to the influence of outer wind irradiation by the inner disk and/or from the effect of the boundary layer on the ionization structure, which are not yet considered in the models.

By analyzing the models for low orbital inclination systems, we found that the synthetic line profiles do not adequately reproduce the line profiles. However, for such cases we found that the profiles are sensitive to the wind geometry and vertical structure. The vertical structure is affected by the gas expansion and geometric radiative dilution that are not taken into account in the plane-parallel approximation. In addition, the interaction among neighboring atmospheres must be taken into account.  This may be crucial for CVs --- here we tried to solve a 3D (or at least 2.5D) problem using a set of 1D models.  Nevertheless, the models  reproduce to reasonable accuracy the high orbital inclination system data. This  leads us to the conclusion that the line profiles of high inclination systems are less dependent on the geometry of the wind, being dictated mostly by the accretion disk parameters.	

In previous works the attention often focussed on \ionciv, because of its strong sensitivity to wind parameters. Nevertheless, it was shown (e.g., \cite{long02} and \cite{long06}) that is an arduous problem to reproduce other UV lines with the same wind parameters. In the present models we attempt to fit simultaneously several line intensities and their profiles. Our results suggest that a highly structured wind can help to improve the fit to the line profiles and the intensity ratios. In addition, we found that it is important to have a 2D consistent structure that take into account the influence of disk radiation on the whole wind. We found model absorption profiles similar to  those observed in the $FUSE$ spectral region \citep{long06,froning03}, but with blue shift velocities higher than observed. This may suggest that these lines come from the photosphere wind transition region, but the wind is less accelerated than calculated here.

In a recent work, \cite{noebauer10} uses a 3D Monte Carlo method developed by \cite{long02}. They show more complex ionization structures, where \ionn{C}{iv}  dominates large extensions of the wind, which bears a stronger emission in the \ionciv\ line. We also found an extended \ionn{C}{iv} region in our winds, but highly concentrated close to the disk axis. 

\begin{figure}
\centering
\includegraphics[height=0.45\columnwidth,width=0.45\columnwidth]{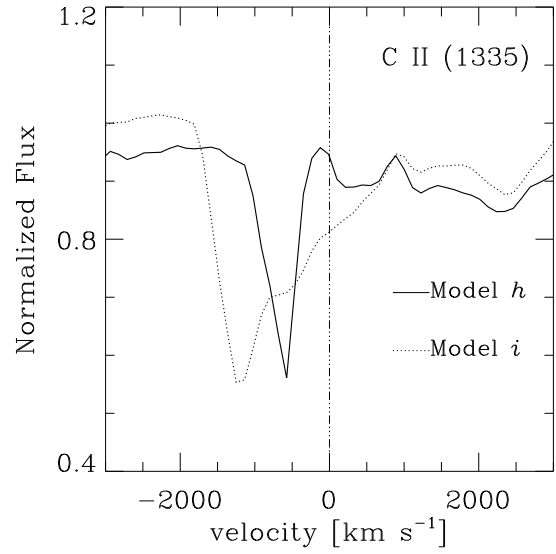} 
\caption{\ioncii\ line profiles for models $h$ and $i$ and an orbital inclination of 30$^\circ$. They show the effect of the different acceleration laws on the photosphere-wind interface region.}\label{fgcii}
\end{figure}

On the other hand, our method is capable of take account consistently the photosphere-wind transition region, although for that we had to give up the 2.5 D description for the wind temperature. We have started a brief analysis of the effects of that region on the emitted spectrum. A diagnostic feature of such regions are the blueshifted absorption components found in low orbital inclination models. We tried to analyze the effect of that interface region on the emitted spectrum using the \ioncii\ line. This is a photospheric line or it is produced in the wind close to disk \citep{noebauer10}. We used two models with different acceleration laws and an orbital inclination $i$=30$^\circ$ (models $h$ and $i$ from table \ref{tbmodels}) in order to test the effect on the profiles. The figure \ref{fgcii} shows that even considering that the line is produced very close to disk photosphere, the acceleration law has a strong influence on the profile and on its equivalent width. These facts show that the interface region structure is important to analyze the spectrum.

Another obstacle in the comparison of disk wind models with observations is the variability seen in the UV resonance lines. Strong variability, not related to the orbital phase, was already found in CV wind lines. \cite{prinja00a} showed that the nova-like BZ Cam has rapid line variability, on timescales as short as minutes suggesting the existence of perturbations that could travel through the wind on that time scale. Other systems show variability in different time scales, for instance V603 Aql \citep{prinja00b} and RW Sex \citep{prinja03}. On the other hand, \cite{hartley02} found variability patterns in the high accretion rate systems IX Vel and V3885 Sgr that could challenge the line-driven hypothesis for those winds.
 
The wind emission has been pointed out as a candidate for solution to the known the color-magnitude problem seen in previous models \citep[e.g.][]{wade88,knigge98}. We found a slight reddening of the spectrum due to the wind which is stronger for higher mass-loss rates and $\lambda$$<$$\text{1500 \AA}$. This is due to the height where the continuum becomes transparent, which is slightly higher when the mass-loss rate is increased, and the corresponding temperature is lower. However, this reddening is not sufficient to fully overcome the UV continuum emission problem. \cite{puebla07} found an average color mismatch of $\sim$0.3 mag between 1500-3000 \AA. Our dense model lowers that value to $\sim$0.24 mag when the 1000-3000 \AA\ region is taken into account.

It is important to summarize the main limitations and the scope of our approximations. The strongest limitation is not taking into account the radiation of the whole disk on the local wind structure (temperature, electron density, population levels and radiation itself). This is more significant for the lines that are formed high above the disk. This point will be improved in future work by adding one dimension to the structure calculus. 

The outer (cooler) atmosphere is mostly responsible for the low ionization emission found in our models that are not observed in low inclination systems. Again, in our approximation a possible explication is that the outer and higher wind regions do not see the radiation that is coming from the hotter inner disk (and boundary layer as well). Hence, a large low ionization volume dominates the outer regions of the wind. 

We arbitrarily included regions with increased wind density trying to improve the line fitting. This also explore the possibility of more complex structures that could produce the line emission. This is not intended to be a definitive or unique model and further exploration is needed.

However, our synthetic spectra present several trends that can be contrasted with observations, for example: i) the dependence of the high-to-low ionization line ratios with the wind temperature. ii) the dependence of the depth of the absorption components with the WD mass. iii) the emission line strength as a function of the disk temperature. iv) our models are able to reproduce the observed rise in the emission lines with orbital inclination and increasing mass-loss rates. The UV FOS and IUE data for the high inclination systems LX Ser, V348 Pup and V347 Pup (systems with different disk temperature) show some of these trends: hotter disks produce stronger lines, and the line ratio \ionn{C}{iv}/\ionn{Si}{iv} falls for cooler disks.

As was already pointed out, a new ingredient of the models is the photosphere wind interface. The effect of such a region on the synthetic spectrum is evident mainly in the “photospheric” lines (e.g. \ioncii) and in the blushifted absorption features. In the later case, however, this effect can be overshadowed by the absorption in the upper regions of the wind. The photosphere-wind interface is insensitive to the of radiation produced at distant regions in the disk. 

The next step to improve our models is to abandon the cylindrical geometry generated by the 1D atmosphere structures, trying to use a hydrodynamically computed geometry to reduce the numbers of free parameters. This alone may not solve all the problems, but will facilitate the model constraining itself. In addition, a method to include the influence of the whole disk radiation on the wind structure will be developed, possibly with the help of Monte-Carlo techniques.

\section{SUMMARY AND CONCLUSIONS}\label{conclusions}

In this work we made an attempt to understand the UV accretion disk and its wind emission. We use a simple 2.5D model formed by a set of 1D models that treat disk-wind interface self-consistently. In general, the geometry of CV systems  does not allow a one-dimensional treatment of the problem. However, in order to make the problem tractable, we constructed a grid of 1D models based on local disk properties, and then combined these models into a 2.5D model. Disk rotation was taken into account --- this is essential  for computing line profiles.

The density and velocity structure of the wind above the disk was determined using line-driven wind theory. Due to the dependence of gravity on height,
the wind acceleration laws differ from stellar results. Both the atmospheric structure and radiative transfer depend on the wind acceleration. We found that the ionization structure and wind temperature strongly depend on the vertical velocity. The Euler equation solutions yield strongly accelerated winds, with maximum acceleration close to the photosphere. Because of this, the atmospheric structure shows strong variations in the photosphere-wind transition region. Higher in the wind, the physical structure is almost constant with $z$, and the ionization state is close to that corresponding to a stellar atmosphere with temperature $T_{eff}(r)$.

From an analysis of synthetic spectra, we find that the wind spectral features are strongly affected by the classical disk properties. The disk temperature sets the wind ionization structure in the sense of the size and location of line formation regions. The primary mass affects the strength of absorption profiles as well as the line emission volume. An increase in the mass loss rate (and wind density), as expected, strengthens the wind characteristics of profiles; the absorption components are deeper and the emission structures are enhanced. An increase in the mass-loss rate also makes the continuum slightly redder especially for $\lambda$$<$1500 \AA. However, this reddening is not enough to overcome the color-magnitude problem which remains present in our models. In agreement with observations, absorption and P~Cygni lines are synthesized for low inclination disks while an emission line spectrum is found for high inclination disks. 

While many characteristics of CV spectra can be explained by our simplified models, discrepancies remain. In future modeling it will be important, for example, to take into account the interaction between different regions of the disk wind. In order to reproduce the UV line intensity ratios observed in high inclination CVs, and to explain variability, condensations in the wind may be necessary.

\acknowledgments

We thank FAPESP (process: 2005/04128-5) for financial support. MPD also acknowledges support from CNPq under grant \#305725. DJH acknowledges partial support by the National Aeronautics and Space Administration under grant ATP 03-0104-0144. We also thank the anonymous referee for his valuable comments that helped us to improve this manuscript.


\clearpage

\clearpage

\begin{deluxetable}{cccccccccccc}
\tabletypesize{\footnotesize}
\tablecaption{Model  Parameters\label{tbmodels}}
\tablewidth{0pt}
\tablehead{
\colhead{Model} &\colhead{\mwd} & \colhead{\rwd} & \colhead{\.{M}$_a$} & \.{M$_w$} & \colhead{T$_{disk}$\tablenotemark{1}}& \colhead{V$_\infty$\tablenotemark{2}} & \colhead{V$_\infty$\tablenotemark{3}} & \colhead{$r_f$\tablenotemark{4}} & \colhead{$\alpha$} & 
\colhead{$k$} & \colhead{$\zeta(x)$}\\
\colhead{} & \colhead{M$_\sun$} & \colhead{10$^{8}$ cm} & \colhead{M$_\sun$ yr $^{-1}$} & \colhead{M$_\sun$ yr $^{-1}$} & \colhead{10$^4$ K} & \colhead{km s$^{-1}$} & \colhead{km s$^{-1}$} & \colhead{10$^9$ cm} & \colhead{} & \colhead{} & \colhead{}
}
\startdata
$a$ & 1.0 & 5.5 & 1.0$\times$10$^{-8}$  & 1.1$\times$10$^{-10}$ & 8.60 & 1540\tablenotemark{5} & 491  & 13.9  & 0.9 & 0.6 & $\zeta_1$ \\
$b$ & 1.0 & 5.5 & 5.0$\times$10$^{-9}$  & 4.0$\times$10$^{-11}$ & 7.20 & 1640\tablenotemark{5} & 461  & 9.0  & 0.9 & 0.6 & $\zeta_1$ \\
$c$ & 1.0 & 5.5 & 1.0$\times$10$^{-9}$  & 7.0$\times$10$^{-12}$ & 4.80 & 1470\tablenotemark{5} & 580  & 6.2 & 0.9 & 0.6 & $\zeta_1$ \\
$d$ & 1.0 & 5.5 & 5.0$\times$10$^{-10}$ & 5.0$\times$10$^{-12}$ & 4.10 & 1110\tablenotemark{5} & 570  & 4.2 & 0.9 & 0.6 & $\zeta_1$ \\
$e$ & 0.8 & 6.7 & 1.0$\times$10$^{-8}$  & 9.3$\times$10$^{-11}$ & 6.80 & 6200\tablenotemark{5} & 1200 & 13.9  & 0.7 & 0.7 & $\zeta_2$ \\
$f$ & 0.8 & 6.7 & 1.0$\times$10$^{-8}$  & 4.7$\times$10$^{-10}$ & 6.80 & 5800\tablenotemark{6} & 908  & 13.9  & 0.5 & 0.9 & $\zeta_2$ \\
$g$ & 0.8 & 6.7 & 1.0$\times$10$^{-9}$  & 1.7$\times$10$^{-12}$ & 3.86 & 1248\tablenotemark{5} & 324  & 4.9 & 0.9 & 0.6 & $\zeta_1$ \\
$h$ & 0.6 & 8.6 & 5.0$\times$10$^{-9}$  & 4.0$\times$10$^{-11}$ & 4.63 & 1182\tablenotemark{5} & 523  & 9.7 & 0.9 & 0.6 & $\zeta_1$ \\
$i$ & 0.6 & 8.6 & 5.0$\times$10$^{-9}$  & 6.5$\times$10$^{-11}$ & 4.63 & 4500\tablenotemark{6} & 1360 & 9.7 & 0.5 & 0.9 & $\zeta_2$
\enddata
\tablenotetext{1}{\begin{footnotesize}Maximum disk effective temperature (Eq. \ref{eqtemp}).\end{footnotesize}}
\tablenotetext{2}{\begin{footnotesize}Maximum vertical velocity V$_{z_\infty}$ (Eq. \ref{eqeuler}) for the first ring.\end{footnotesize}}
\tablenotetext{3}{\begin{footnotesize}Maximum vertical velocity V$_{z_\infty}$ calculated with Eq. \ref{eqeuler} for the last ring at $r_f$.\end{footnotesize}}
\tablenotetext{4}{\begin{footnotesize}External wind atmosphere radius, T($r_f$)$\sim$14000 K. \end{footnotesize}}
\tablenotetext{5}{\begin{footnotesize}Lower branch solutions.\end{footnotesize}}
\tablenotetext{6}{\begin{footnotesize}Upper branch solutions.\end{footnotesize}}
\end{deluxetable}

\begin{deluxetable}{cc}
\tabletypesize{\normalsize}
\tablecaption{Geometrical Parameters\label{tbgeopar}}
\tablewidth{0pt}
\tablehead{\colhead{Parameter} & \colhead{Values} 
}
\startdata
$i$ ($^\circ$) & 30, 40, 50, 60, 70, 80   \\
$r_{i}$/\rwd &  1.05 \\
$\theta_1$ ($^\circ$) & 0, 5, 15 \\
$\theta_2$ ($^\circ$) & 0, 15, 45  \\
Z$_{max}$/R$_{disk}$ &  1.0 \\
$r_{C}$/$r_{f}$ & 1.0
\enddata
\end{deluxetable}

\begin{deluxetable}{ccccccc}
\tabletypesize{\footnotesize}
\tablecaption{Physical characteristics of each ring used to calculate the model ``$e$'', namely: the radius of each ring, the mass loss flux, the terminal velocity and effective temperature. The star-like wind parameters are also listed. \label{tbmodele}}
\tablewidth{0pt}
\tablehead{ & \multicolumn{4}{c}{Disk parameters} & \multicolumn{2}{c}{Star-like parameters} \\
\colhead{Ring} & \colhead{r} & \colhead{\.{m}(r)} & \colhead{$v_\infty$} & \colhead{T$_{eff}$} & \colhead{L} & \colhead{\.{M}$_v$}\\
& \colhead{\rwd} & \colhead{gr cm$^{-2}$ s$^{-1}$} & \colhead{km s$^{-1}$} & \colhead{10$^4$ K} &   \colhead{L$_\odot$} & \colhead{M$_\odot$ yr$^{-1}$}
}
\startdata
1 & 1.05 & 2.6$\times$10$^{-5}$ & 4043 & 5.42   &7.9$\times$10$^{5}$&1.2$\times$10$^{-6}$\\
2 & 1.52 & 1.4$\times$10$^{-4}$ & 3351 & 6.87   &2.0$\times$10$^{6}$&5.5$\times$10$^{-6}$\\
3 & 2.20 & 9.6$\times$10$^{-5}$ & 2770 & 5.96   &1.1$\times$10$^{6}$&3.6$\times$10$^{-6}$\\
4 & 3.19 & 4.6$\times$10$^{-5}$ & 2500 & 4.86   &5.1$\times$10$^{5}$&1.7$\times$10$^{-6}$\\
5 & 4.63 & 1.9$\times$10$^{-5}$ & 2500 & 3.86   &2.0$\times$10$^{5}$&8.3$\times$10$^{-7}$\\
6 & 6.70 & 7.3$\times$10$^{-6}$ & 2400 & 3.03   &7.7$\times$10$^{4}$&3.2$\times$10$^{-7}$\\
7 & 9.72 & 2.7$\times$10$^{-6}$ & 1950 & 2.35   &2.7$\times$10$^{4}$&1.2$\times$10$^{-7}$\\
8 & 14.08 & 9.6$\times$10$^{-7}$ & 1600 & 1.81   &9.9$\times$10$^{3}$&9.3$\times$10$^{-8}$\\
9 & 20.41 & 3.3$\times$10$^{-7}$ & 1260 & 1.39   &3.4$\times$10$^{3}$&1.4$\times$10$^{-8}$\\
10 & 29.58 & 1.2$\times$10$^{-7}$ & 980 & 1.06   &1.19$\times$10$^{3}$&1.1$\times$10$^{-8}$\\
11 & 42.80 & - & - & 0.81  & - & -\\
12 & 62.10 & - & - & 0.62  & - & -
\enddata
\end{deluxetable}

\begin{deluxetable}{lcclcc}
\tabletypesize{\footnotesize}
\tablecaption{Ionization states used to calculate the atmospheric structures.\label{tbions}}
\tablewidth{0pt}
\tablehead{ \colhead{Ion} & \colhead{Levels} & \colhead{Super-levels} & \colhead{Ion} & \colhead{Levels} & \colhead{Super-Levels}
}
\startdata
H I & 20 & 20 & O V & 41 & 41\\
He I & 40 & 40 & O VI & 13 & 13 \\
He II & 22 & 22 & O VII & 1 & 1 \\
C II& 0 & 14 &  Ne II & 0 & 14  \\
C III& 62 & 62 & Ne III & 23 & 13 \\
CIV & 13 & 13 & Ne IV & 17 & 17 \\
N II & 36 & 36 & Si III & 20 & 20 \\
N III & 34 & 34 & Si IV & 22 & 22 \\
N IV & 90 & 90 & Fe II\tablenotemark{1} & 135 & 135 \\
N V & 33 & 33 & Fe III\tablenotemark{1} & 96 & 96 \\
N VI & 1 & 1 &  Fe IV & 100 & 100 \\
O II & 46 & 46 & Fe V & 139 & 139 \\
O III & 92 & 92 & Fe VI & 44 & 44 \\
O IV & 29 & 29 & Fe VII & 29 & 29
\enddata
\tablenotetext{1}{\begin{footnotesize} Ions used when T$_{eff} \lesssim$ 20000 K. \end{footnotesize}}
\end{deluxetable}

\begin{deluxetable}{lccccccc}
\tablecaption{System Parameters\label{tbsyspar}}
\tablewidth{0pt}
\tablehead{
\colhead{System} & \colhead{P$_{orb}$} & \colhead{\mwd} & \colhead{M$_2$} & \colhead{d} &\colhead{\.{M}$_a$} & \colhead{$E(B-V)$} & \colhead{References} \\
\colhead{} & \colhead{days} & \colhead{M$_\sun$} & \colhead{M$_\sun$} & \colhead{pc} & \colhead{\msun\ yr$^{-1}$} & \colhead{} & \colhead{ }
}
\startdata
RW Tri & 0.25 & 0.55 & 0.35 &310-380 & \sci{4.6}{-9} & 0.1 & 1,2,3,4\\
V347 Pup & 0.23 & 0.63 & 0.55 & 510 & \sci{6.0}{-9} & 0.06 & 1,4,5,6 
\enddata
\tablerefs{ (1) \citealt{ritter03}, (2) \citealt{poole03}, (3) \citealt{mcarthur99}, (4) \citealt{puebla07}, (5) \citealt{thoroughgood05}, (6) \citealt{diaz99}.}
\end{deluxetable}

\begin{deluxetable}{lcccc}
\tablecaption{Models Parameters\label{tbmodpar}}
\tablewidth{0pt}
\tablehead{
\colhead{} & \multicolumn{2}{c}{RW Tri} & \multicolumn{2}{c}{V347 Pup}\\
\colhead{Parameter} & \colhead{$b'$} & \colhead{$h'$} & \colhead{$b''$} & \colhead{$h''$}
}
\startdata
P$_{orb}$ (days) & 0.3 & 0.3 & 0.3 & 0.3 \\ 
\mwd (\msun) & 1.0 & 0.6 & 1.0 & 0.6 \\
\.{M}$_a$ (\sunyr) & \sci{5}{-9} & \sci{5}{-9} & \sci{5}{-9} & \sci{5}{-9} \\
\.{M}$_w$ (\sunyr) & \sci{8.2}{-11} & \sci{1.1}{-10} & \sci{8.2}{-11} & \sci{8.1}{-11} \\
$i$ ($^\circ$) & 70 & 70 & 80 & 80 \\
$R_{disk}$ (10$^9$ cm) & 45.9 & 39.7 & 45.9 & 39.7 \\ 
$r_i$ (10$^9$ cm) & 0.58 & 0.84 & 0.58 & 0.84 \\ 
$r_f$ (10$^9$ cm) & 9.05 & 9.74 & 9.05 & 9.74 \\
$r_C$ (10$^9$ cm) & 1.87 & 9.74 & 9.05 & 9.74 \\
$\theta_1$ ($^\circ$) & 5 & 2 & 5 & 5 \\
$\theta_2$ ($^\circ$) & 35 & 10 & 45 & 45 \\
$r_{d_1}-r_{d_2}$ (10$^9$ cm)\tablenotemark{1}  & 0.83-1.25 & 1.25-3.47 & 0.83-1.25 & 1.25-1.73 \\
\enddata
\tablenotetext{1}{\begin{footnotesize} Radial region of enhanced wind density. See text. \end{footnotesize}}
\end{deluxetable}

\begin{deluxetable}{lccccc}
\tablecaption{Equivalent widths (EW) for two RW Tri models.\label{tbrwtri}}
\tablewidth{0pt}
\tablehead{
\colhead{Line} & \colhead{EW$_{M}$\tablenotemark{1}} & \colhead{EW$_{b'}$} & \colhead{EW$_{h'}$} & \multicolumn{2}{c}{Profile Match} \\
\colhead{} & \colhead{(\AA)} & \colhead{(\AA)} & \colhead{(\AA)} & \colhead{Model $b'$\tablenotemark{2}} & \colhead{Model $h'$\tablenotemark{2}} 
}
\startdata
\ionn{C}{iii} & 12.3 & 18.3 & 15.5 & good & good \\
\ionn{N}{v} & 25.5 & 43.6 & 57.0 & medium &  bad \\
\ionn{Si}{iv} & 28.7 & 29.0 & 28.0 & good &  good \\
\ionn{C}{iv} & 75.0 & 73.0 & 53.6 & good &  medium  
\enddata
\tablenotetext{1}{\begin{footnotesize}Measured equivalent width from RW Tri UV data.\end{footnotesize}}
\tablenotetext{2}{\begin{footnotesize}Modified models from Table \ref{tbmodels} standard parameters (see text).\end{footnotesize}}
\end{deluxetable}

\begin{deluxetable}{lccccc}
\tablecaption{Equivalent widths (EW) for two V347 Pup models.\label{tbv347pup}}
\tablewidth{0pt}
\tablehead{
\colhead{Line} & \colhead{EW$_{M}$\tablenotemark{1}} & \colhead{EW$_{b''}$} & \colhead{EW$_{h''}$} & \multicolumn{2}{c}{Profile Match} \\
\colhead{} & \colhead{(\AA)} & \colhead{(\AA)} & \colhead{(\AA)} & \colhead{Model $b''$\tablenotemark{2}} & \colhead{Model $h''$\tablenotemark{2}}
}
\startdata
\ionn{C}{iii} & 50.0 & 70.0 & 57.0 & medium & medium \\
\ionn{N}{v} & 167.0 & 76.0 & 62.0 & bad &  bad\\
\ionn{Si}{iv} & 127.0 & 137.0 & 110.0 & medium & good\\
\ionn{C}{iv} & 244.0 & 134.0 & 121.0 & bad &  bad  
\enddata
\tablenotetext{1}{\begin{footnotesize}Measured equivalent width from V347 Pup UV data.\end{footnotesize}}
\tablenotetext{2}{\begin{footnotesize}Modified models from Table \ref{tbmodels} standard parameters (see text).\end{footnotesize}}
\end{deluxetable}

\end{document}